\documentclass[preprint2]{aastex}
\usepackage{apjfonts}
\slugcomment{}
\shorttitle{Jet and Wind Driven Outflows in NGC 3079}
\shortauthors{Cecil et al.}

\begin{document}

\title{Jet- and Wind-Driven Ionized Outflows in the Superbubble and Star-Forming
Disk of NGC 3079\footnote{%
Based on observations made with the NASA/ESA \emph{Hubble Space Telescope,}
obtained at the Space Telescope Science Institute, which is operated by the
Association of Universities for Research in Astronomy, Inc., under NASA contract
NAS5-26555. These observations are associated with proposal ID GO-6674.
}}

\author{Gerald Cecil\altaffilmark{2}}
\altaffiltext{2}{SOAR Telescope Project, NOAO, Tucson AZ 85726-6732}

\affil{Dept. of Physics \& Astronomy, University of N. Carolina, Chapel Hill, NC 27599-3255}

\email{gerald@thececils.org}

\author{Joss Bland-Hawthorn}

\affil{Anglo-Australian Observatory, Epping, NSW, Australia}

\email{jbh@aao.gov.au}

\author{Sylvain Veilleux\footnote{%
Cottrell Scholar of the Research Corporation
}}

\affil{Dept. of Astronomy, University of Maryland, College Park, MD 20742}

\email{veilleux@astro.umd.edu}

\and{}

\author{Alexei V. Filippenko}

\affil{Dept. of Astronomy, University of California, Berkeley, CA 94720-3411}

\email{alex@astro.berkeley.edu}

\begin{abstract}
\emph{HST} WFPC2 images are presented that span the inner $\sim$19 kpc diameter of
the edge-on galaxy NGC 3079; they are combined with optical, emission-line imaging
spectrophotometry and VLA images of radio polarization vectors and rotation measures.
Ionized gas filaments within 9-kpc diameter project up to 3 kpc above the disk,
with the brightest forming the $ \approx 1 $ kpc diameter superbubble. They
are often resolved into strands $ \approx 0\farcs 3 $ (25 pc) wide, which
emerge from the nuclear CO ring as five distinct streams with
large velocities and velocity dispersions (FWHM $ \approx  $450 km~s$ ^{-1} $).
The brightest
stream emits $\approx 10\%$ of the superbubble H$\alpha$ flux and extends
for 250 pc along the axis of the VLBI radio jet to one corner of the
base of the superbubble.
The other four streams are not connected to the jet, instead curving up to
the vertical $ \approx 0.6 $ kpc above the galaxy disk, then dispersing as a
spray of
droplets each with $ \approx 10^{3}\sqrt{f}M_{\bigodot } $ of ionized
gas (the volume filling factor $f>3\times 10^{-3} $).  
Shredded clumps of disk gas form a similar structure in hydrodynamical 
models of a galaxy-scale wind. The
pattern of magnetic fields and the gaseous kinematics
also suggest that a wind of mechanical luminosity $ L_{w}\approx 10^{43} $
ergs~s$ ^{-1} $ has stagnated in the galaxy disk at a radius of $ \sim 800 $
pc, has flared to larger radii with increasing height as the balancing
ISM pressure reduces above the disk, and has entrained dense clouds 
into a {}``mushroom
vortex{}'' above the disk. H$ \alpha  $ emissivity of the filaments limits
densities to $ n_{e}>4.3f^{-1/2}\,  $cm$ ^{-3} $, hence kinetic energy
and momentum to $ (0.4-5)\times 10^{55}\sqrt{f} $ ergs and $ (1.6-6)\times 10^{47}\sqrt{f} $
dyne~s, respectively; the ranges result from uncertain space velocities.
A prominent star-forming complex elsewhere in the galaxy
shows a striking spray of linear filaments that extend for hundreds of parsecs
to end in unresolved {}``bullets.{}''
\end{abstract}

\keywords{galaxies: individual (NGC 3079) --- galaxies: active --- galaxies: jets ---
galaxies: kinematics and dynamics --- galaxies: ISM --- galaxies: magnetic fields}

\section{INTRODUCTION}

An energetic outflow can inflate a {}``superbubble{}'' to a size that
exceeds the scale height of the interstellar medium (ISM). If the bubble
ruptures, ejecta can alter chemical abundances across the galaxy. The superbubble
is comprised of various shock fronts in the ISM and the outflow ejecta. The ISM
and ejecta have several phases with different density, temperature, embedded magnetic
fields, and dust distributions, and the shocks have different properties depending
upon the phase through which they are propagating. The flow must therefore
be observed across many wavebands from the radio to hard X-rays, but is 
often seen most clearly by imaging spectrophotometry
of optical emission lines. Spatial resolutions and sensitivities of
current instruments suffice to diagnose gaseous conditions for only the nearest
and brightest superbubbles (e.g. M82, Shopbell \& Bland-Hawthorn 1998;
NGC 253, Forbes et al. 2000; NGC 4438, Kenney \& Yale 2000; Kenney
et al. 1995) where the angular scale is large enough to constrain
dynamical models. 

One of the clearest examples is the superbubble emerging from the disk of the
edge-on SB-pec \citep{de91} galaxy NGC 3079; 
galaxy properties are summarized in Table \ref{tab:intro}. The X-shaped
radio structure on the nucleus discovered by \cite{dB77} was subsequently resolved
into a figure-8 pattern by \citet[hereafter DS]{DS88} in the radio continuum, and by \cite{Fo86}
in optical line emission. The nucleus contains
the most luminous known H$ _{2} $O megamaser \citep{He84} which \cite{Tr98}
resolved spatially, and has large extinction, so conditions there 
have been established from IR emission lines and radio-frequency spectra
\citep{Ir88,Ir92,Ba95} which show outflow signatures.

\placetable{tab:intro}

\citet{HAM} and
\citet[hereafter Paper I] {FS92} obtained long-slit spectra of the
ionized gas associated with the superbubble, and found large velocity 
gradients and emission-line widths in the patterns expected for
an expanding structure. In \citet[hereafter Paper II]{Ve94} we used Fabry-Perot
spectrophotometry to map the ionized superbubble, showing
that its pattern of radial
velocities is consistent with an outflow whose space velocities
increase from the nucleus.

Because of its considerable extinction \citep{Pa97,So00}, the nuclear power
source manifests itself primarily by reprocessed radiation.
How the superbubble filaments are powered is thus uncertain.
Their total ionized mass and kinetic energy (KE)
are scaled by gas densities whose values
are poorly constrained due to blending in ground-based spectra (of the red {[}\ion{S}{2}{]}
doublet flux ratio) and images (to establish the emitting volume). DS
modeled the radio structure as a wind-blown bubble from the active galactic
nucleus (AGN), and \cite{Ha95} concluded that an AGN wind is responsible largely
because hot dust characteristic of a starburst is absent; \cite{Is98} 
subsequently found hot dust out to a radius of 300 pc.
The structure appears to
be in a {}``breakout{}'' stage, with the driving wind expanding freely into a
halo of diffuse X-ray emission \citep[hereafter PTV]{Pi98}. 
An AGN in NGC 3079 opens the possibility of a more collimated
outflow, and indeed a jet has been proposed \citep{Ir88,Tr98,Sa00} to explain 
several aligned, VLBI-scale radio knots along with the maser kinematics \citep{Sa00}.
\cite{Ir88} argued
that a jet alone may inflate the superbubble.

If the wind arises from an optically extinguished
AGN, it provides a nearby laboratory
for study of processes thought to have operated in
dust-shrouded proto-quasars. If the wind is associated with a
starburst, the processes operating in IR ultra-luminous, distant galaxies can be
studied in detail.
To address these issues, in \S2 we present
Hubble Space Telescope (\emph{HST}) WFPC2 and unpublished
VLA\footnote{%
The VLA is a facility of the National Radio Astronomy Observatory. The National
Radio Astronomy Observatory is a facility of the National Science Foundation
operated under cooperative agreement by Associated Universities, Inc.
} radio continuum images of the galaxy disk and superbubble.
In \S3 we derive properties of the magnetic fields, and show that ionized
filaments are often resolved transversely so that we can constrain
their gas densities and filling factor from recombination fluxes. 
Some filaments with large velocity dispersions align with
the VLBI-scale axis, extending the directly detected
influence of the jet to 250 pc radius.
In \S4 we discuss the dynamical state and power source of the superbubble. 
We outline some future observations in \S5, then
summarize our results and conclusions in \S6.
\placetable{tab:observations}

\section{OBSERVATIONS AND REDUCTIONS}

\subsection{\label{sec:hstobs}\emph{HST} WFPC2 Images}

For sharpest images in exposures of reasonable duration, we
spanned the inner parts of the galaxy with two chips of the WFC but ``dithered"
the exposures between two points separated by
2\farcs50 (confirmed to $ \pm 0\farcs 02 $ by cross-correlating image pairs);
see Table \ref{tab:observations}.
We imaged the galaxy on two rather than three WFC detectors
to maximize coverage perpendicular to --- rather than along --- the plane of
the galaxy disk. The PC chip covered the outskirts of NGC 3079,
and did not obtain useful data.
Our filters passed the I-band (F814W)
and [\ion{N}{2}]$\lambda\lambda$6548,6583+H$\alpha$ (F658N) emission-line complex.
The two I-band exposures were sufficiently short that we did not need to
obtain multiple images to identify cosmic-ray hits. 
Four narrow-band F658N images were made, two at each dither position.

``Warm pixels" were treated using the appropriate dark frame that became
available several weeks after our exposures.
The dithered frame gave us information at these locations, so
we interpolated linearly each frame
to the position of its partner then patched in the affected pixels. The handful
of residual cosmic rays and warm pixels in the combined dither images were
removed manually.
Because only a single continuum image was obtained at
each dither position, cosmic rays were removed by first interpolating linearly
each frame to the position of its partner, then selecting the minimum pixel
value at each point from this image and its uninterpolated partner. Six counts
were
added to the uninterpolated frame (an offset that was determined empirically)
so that the interpolated frame contributed only at the locations of cosmic rays
or warm pixels in the uninterpolated frame. Less than 0.1\% of the pixels
in each frame were in this state, very few across the superbubble. 

After cleaning artifacts and scaling to a common total exposure, we interleaved
the frames at the two dither positions onto a grid of 0\farcs0707 pixels. 
In the continuum image 90\% of a star's light
is encircled within $ 0\farcs 38 $ diameter ($ 0\farcs 226\pm 0\farcs 003 $
FWHM Moffat function), as expected for a two-point dither on the WFC chips
at this wavelength. We scaled the continuum image fluxes to
ensure a {}``line{}''-only image after subtraction. 
To place the I-band image on
an absolute flux scale, we used the color term for an SBc galaxy
given in Table 6.2 of the WFPC2 Instrument Handbook.\footnote{%
\url{www.stsci.edu/instruments/wfpc2/Wfpc2\_hand/wfpc2\_handbook.html}}

\begin{figure}
\includegraphics[scale=0.75,angle=90]{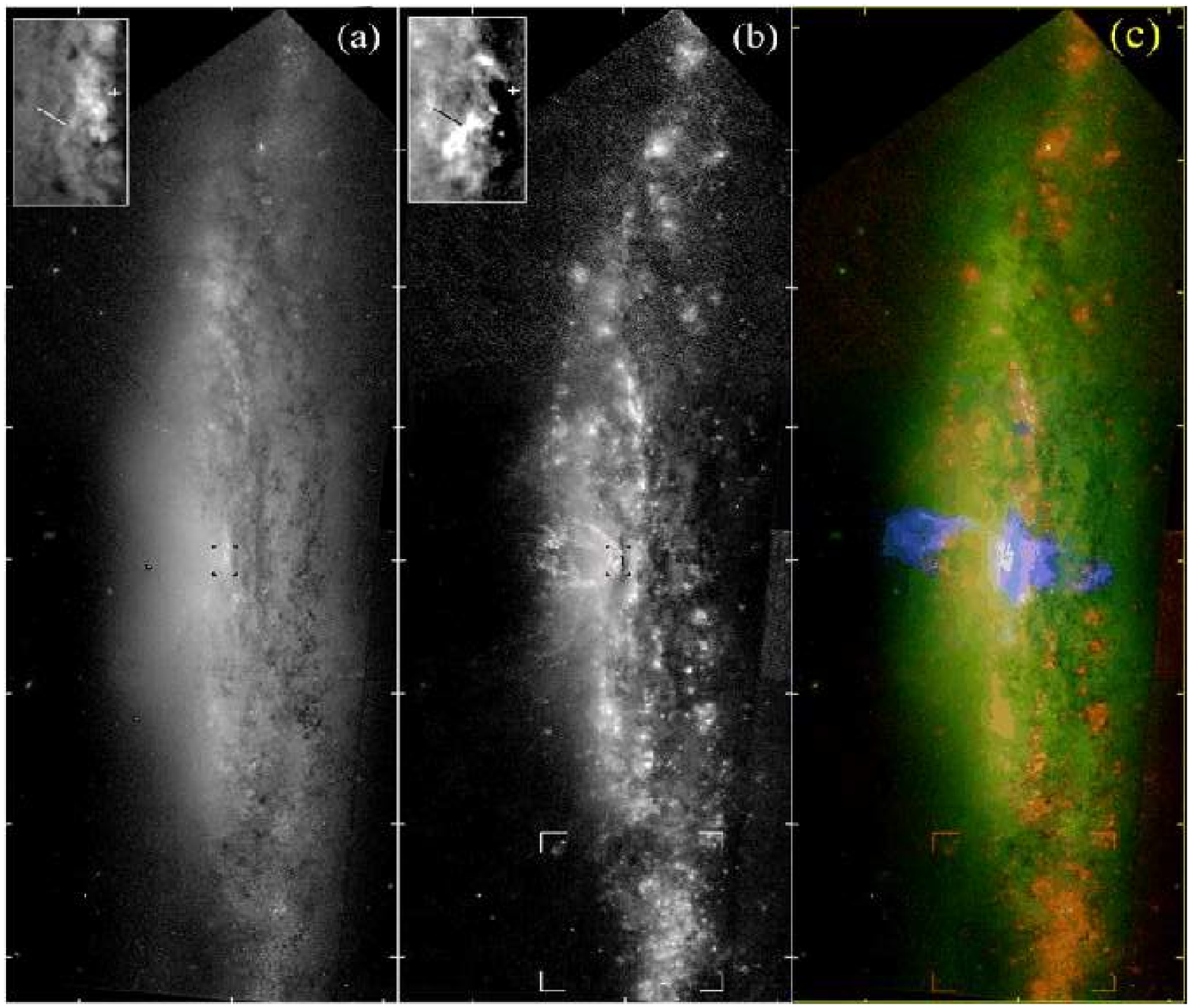}
\caption{\emph{HST} WFPC2 images of NGC 3079, showing the log of intensity, 
with P.A. $-$10\arcdeg\ at top and ticks every 30\arcsec\ = 2.5 kpc.
The top parts are noisier because they come from shorter, archival exposures.
The I-band image (a) has been unsharp-masked to increase
the contrast and prominence of the bilobal pattern of scattered light and
dust filaments in the galaxy bulge. The image of
[\ion{N}{2}]$\lambda$6583+H$\alpha$ line emission (b)
shows narrow ionized filaments that rise above the galaxy disk to envelope
the prominent
superbubble. The region spanned by Fig.\ \ref{fig:blast} is delineated.
The nuclear region is enlarged in the inset panels, with the optically obscured
nucleus marked with a ``+"; the I-band panel was obtained with the PC CCD
of WFPC2, the line image is from dithered WFC CCD frames.
Note the plume marked in both panels, 
which aligns with the VLBI-scale radio jet.
(c) Composite of the
[\ion{N}{2}]$\lambda$6583+H$\alpha$ emission-line 
and I-band images in red and green, respectively, and the VLA 3.8 cm 
continuum in blue.  \label{fig:overall}}
\end{figure}

We obtained additional images from the \emph{HST} archive (P.I.\ H.\ Falcke),
see Table \ref{tab:observations}.
Signal/noise ratios are smaller than those of our images because
the galaxy nucleus was placed on the PC CCD and shorter exposures were used.  
However, the I-band
archive image has slightly higher resolution on the nucleus than our
dithered WFC images, and places the N half of galaxy on one of the WFC
CCD's, complementing our coverage.
Composite images are shown in Fig.\ \ref{fig:overall}.
\placefigure{fig:overall}

\begin{figure}
\includegraphics[scale=0.53]{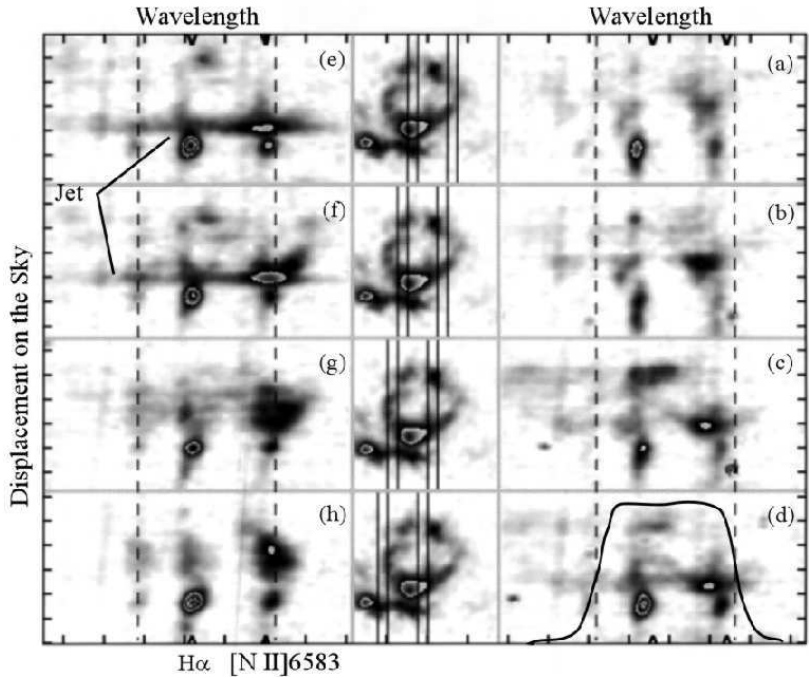}
\caption{
Fabry-Perot spectra of the superbubble (from 
Paper II), summed along the indicated ``long slits", and oriented with
P.A.\ 80\arcdeg\ at top; ticks are every 4\arcsec\ vertically and every
500 km~s$^{-1}$ horizontally. Complex velocities across the
superbubble blend the lines in the {[}\ion{N}{2}{]},H\protect$ \alpha \protect $
complex. The transmission profile of WFPC2
filter F658N is shown at bottom right, with
dashed lines intersecting at 50\% peak transmission. Arrowheads mark
the galaxy 
systemic velocity for the two bright spectral lines. Note broad 
spectral lines at the base of the superbubble, labeled 
``jet" in panels (e) and (f).\label{fig:bubvels}}
\end{figure}

\begin{figure}
\includegraphics[scale=0.7]{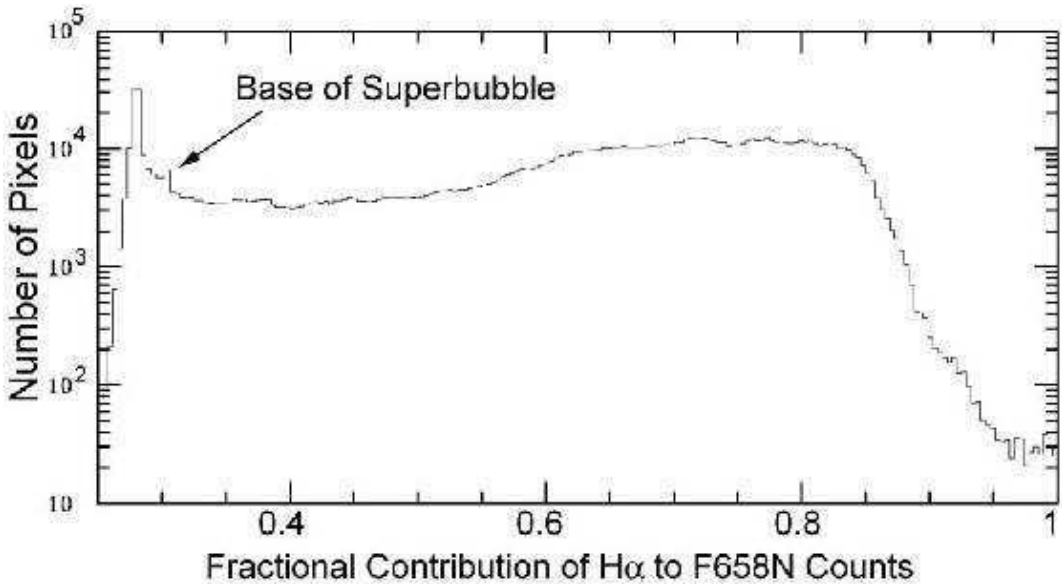}
\caption{
How H\protect$ \alpha \protect $ contributes to the total flux measured
through WFPC2 filter F658N.  \label{fig:weights}}
\end{figure}

In the F658N$ - $F814W image, the flux contributions of the three emission-lines
{[}\ion{N}{2}{]}$ \lambda \lambda  $6548,6583+H$ \alpha  $ passed by the narrower
filter vary across the disk and especially across the superbubble. In \citet[hereafter Paper III]{Ve95}
and \citet[hereafter Paper IV]{Ve99} we presented flux maps and velocity fields,
respectively, for ionized gas outside the superbubble. We showed
in Paper II
that gas in the interval $ 1\la R\la 2.5 $ kpc has average line widths of
$ \approx 150 $ km~s$ ^{-1}, $ dropping to 100 km~s$ ^{-1} $ farther
out. We used these values to reconstruct emission-line profiles of the {[}\ion{N}{2}{]}$ \lambda \lambda  $6548,6583+H$ \alpha  $
complex at ground-based resolution (1\arcsec\ FWHM), and then multiplied the
result by the F658N filter profile to find the average flux fraction contributed
by H$ \alpha  $ at each point. Velocity gradients across the superbubble
are large and vary in a complex fashion, see Fig.\ \ref{fig:bubvels}.
In Paper I we had decomposed the profile of each spectral line of the {[}\ion{N}{2}{]}$ \lambda \lambda 6548,6583+$H$ \alpha  $
complex into 3 Gaussian subsystems within rectangular regions 0\farcs57 high$ \times  $2\farcs3
wide. Using this decomposition, we synthesized the complexes, multiplied them
by the F658N profile, and inserted the resulting H$ \alpha  $ weights into
our weighted image. Fig.\ \ref{fig:weights} shows the distribution of weights.
Corrections exceed those for the disk: at
the base of the bubble only 25\% of the line flux through filter F658N comes
from H$ \alpha  $ (the spike at left in Fig.\ \ref{fig:weights}). 
\placefigure{fig:bubvels}
\placefigure{fig:weights}

\begin{figure}
\includegraphics[scale=0.81]{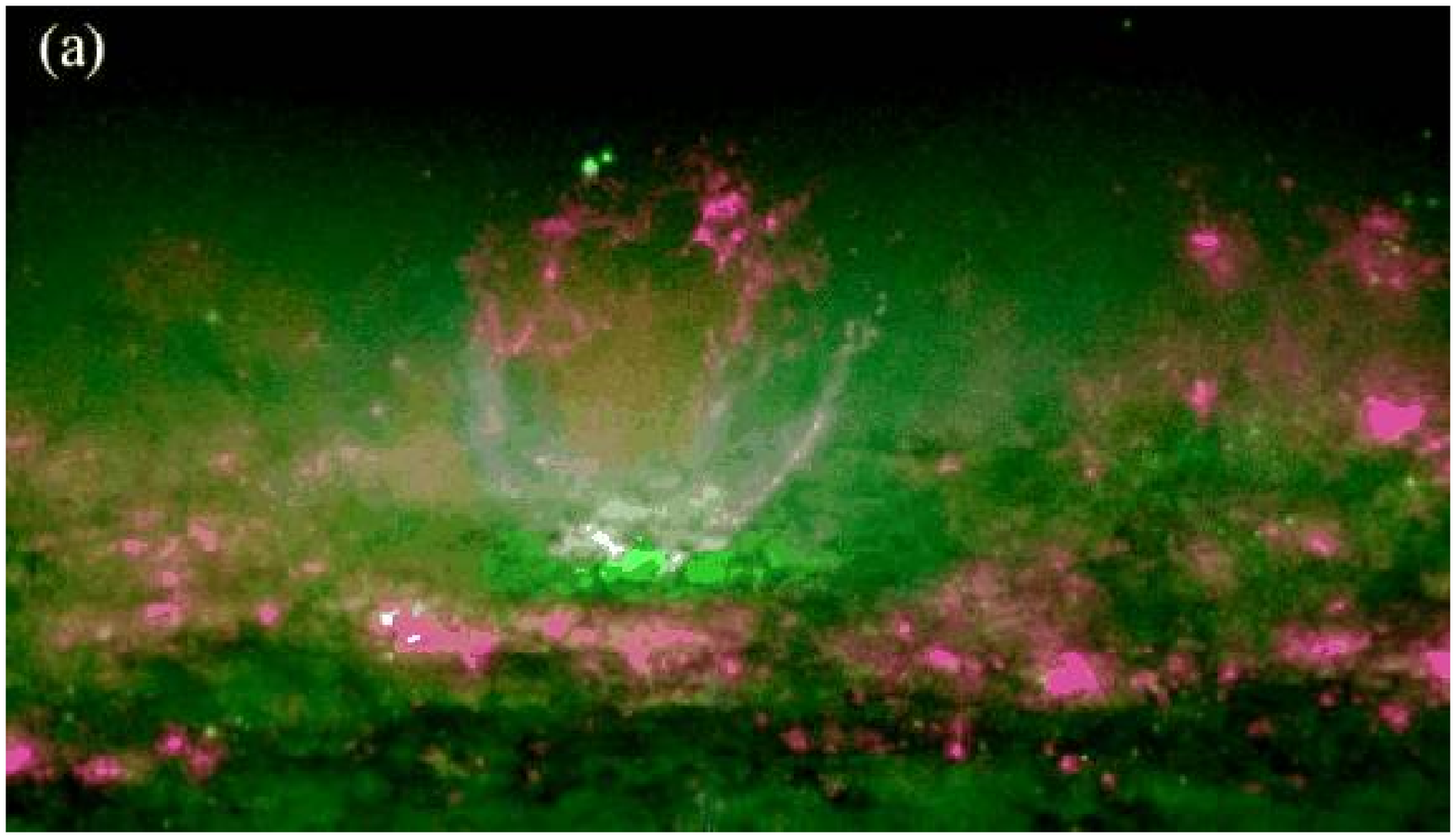}
\includegraphics[scale=0.7]{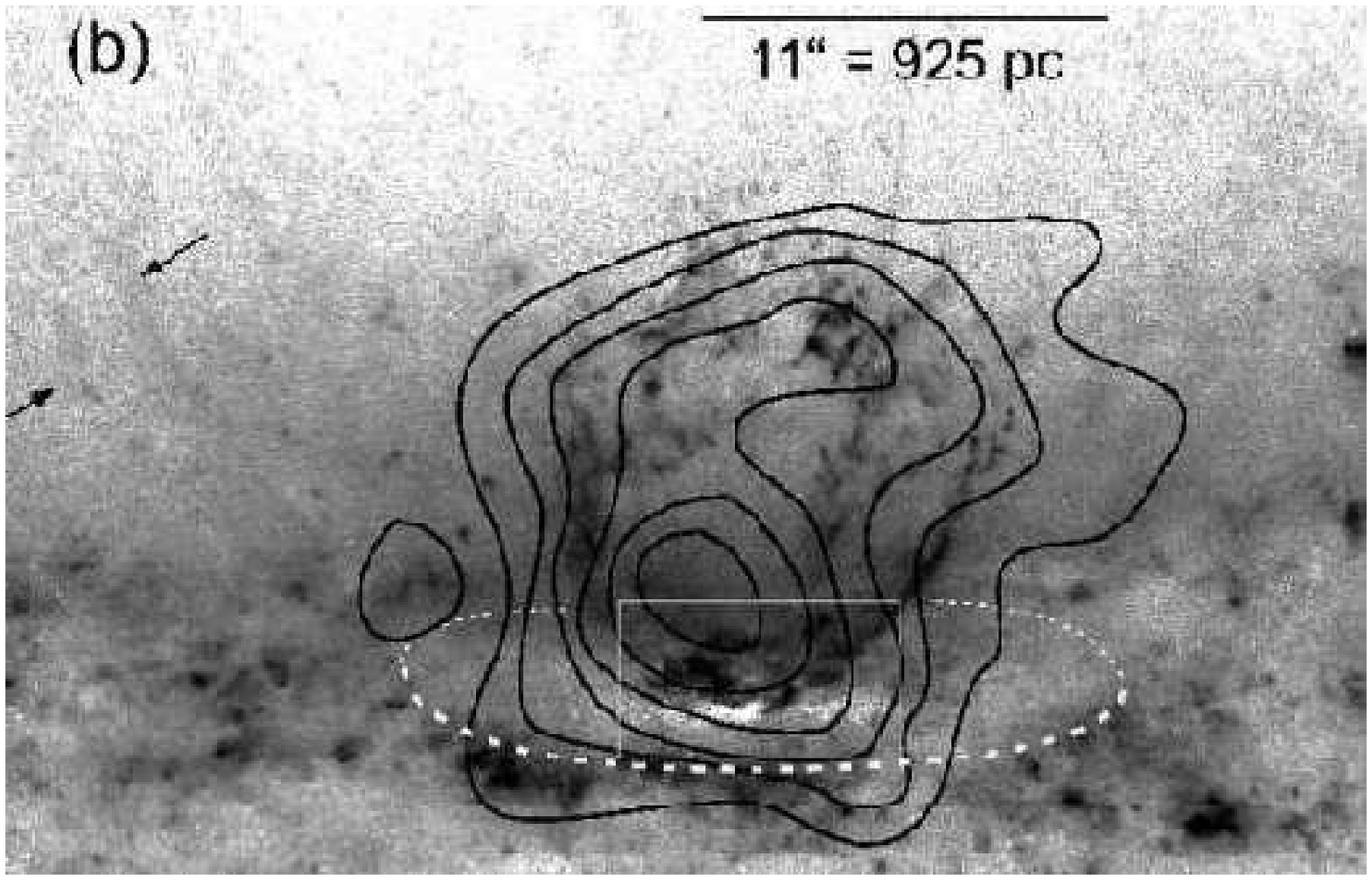}
\caption{\label{fig:bubble}(a) A concave bowl is shown enveloping the superbubble in a region 50\arcsec\ wide;
its base appears to be devoid of H~II regions.
I-band starlight is green, H\protect$ \alpha \protect $ red, and {[}\ion{N}{2}{]}
blue. Note that some dust plumes align with ionized filaments in
the bottom third of the superbubble while others at left and right
arc up and away from
the bowl. The brightest plume at the base of the superbubble aligns with
the VLBI-scale jet \citep{Tr98} (see also the 
inserts in Fig.\ \ref{fig:overall}).
(b) Contours of the \emph{ROSAT} HRI emission (values given in Fig.\ 4 of 
PTV) are shown
atop the \emph{HST} line image, with registration uncertainties of \protect$ \pm 2\arcsec \protect $.
The box delineates the region in Fig. \ref{fig:closeup}a, while the
white dashed ellipse shows the orientation of the galaxy disk.}
\end{figure}

Next, we used the WFPC2 Exposure Calculator\footnote{%
\url{www.stsci.edu/wfpc2/etc.html}
} to find the H$ \alpha  $ fluxes of H II regions whose
{[}\ion{N}{2}{]}$ \lambda  $6583/H$ \alpha  $ flux ratios were
measured in our ground-based data.
From the sum of counts at these regions, we placed
H$\alpha$ fluxes in our interleaved, line-only frame on an absolute scale.
The major change over Fig.\ \ref{fig:overall}b is to darken the lower half of
the bubble where {[}\ion{N}{2}{]}$ \lambda  $6583/H$ \alpha >1 $
(Paper I); see Fig.\ \ref{fig:bubble}a.
\placefigure{fig:bubble}

\subsection{VLA Flux, Polarization, and Rotation Measure Images}

\begin{figure}
\includegraphics[scale=0.83]{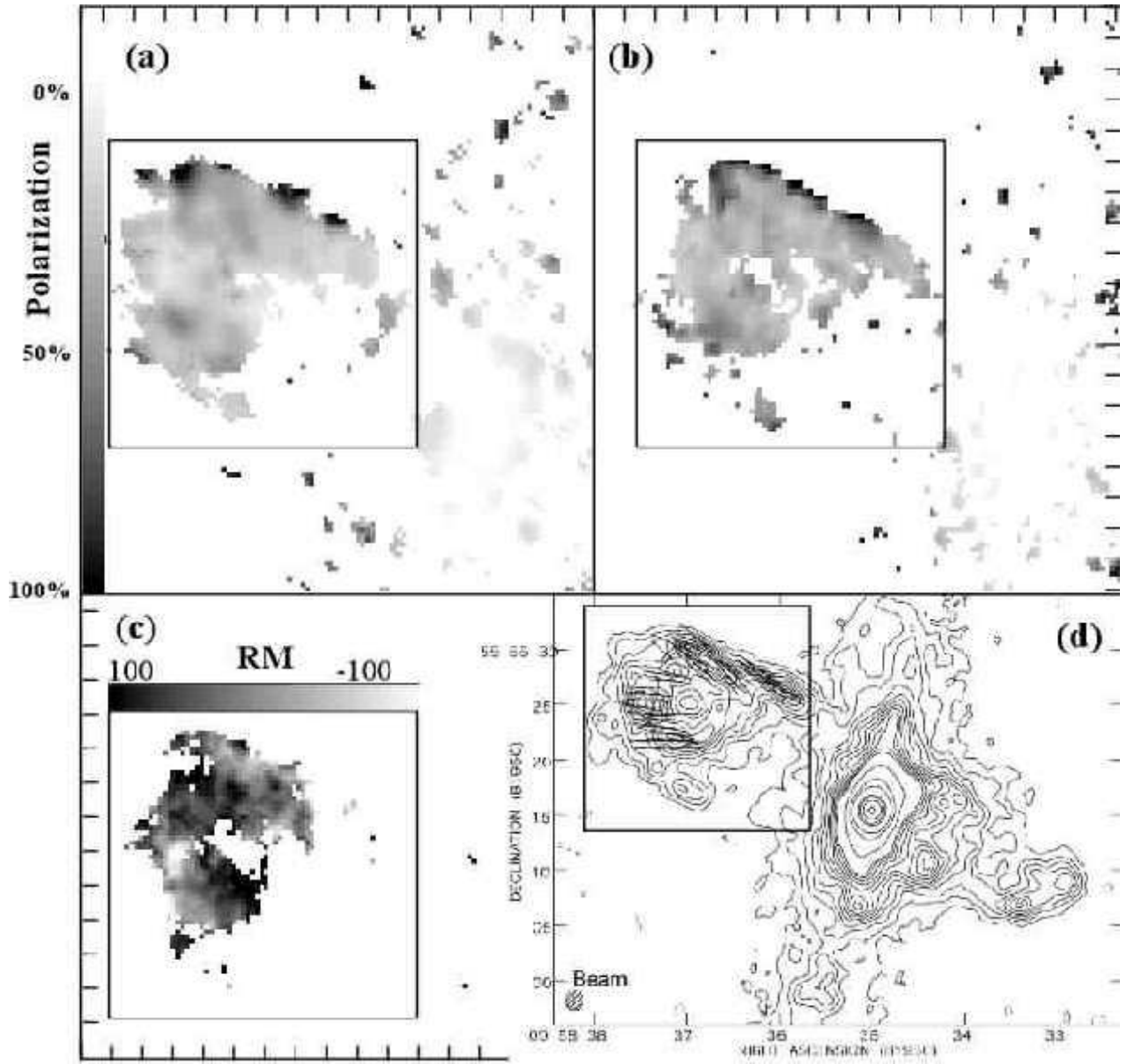}
\caption{
(a) Polarized 6 cm flux is shown as an inverted greyscale. (b)
Same as (a) now for polarized 3.8 cm flux. (c) Derived rotation measures, greyscale
spans $\pm100$ radians~m\protect$ ^{-2}.\protect $ (d) The box delineates the
same region as in the other panels. Contours are 
at levels (90, 50, 30, 20, 5, 2.5, 1, 0.6, 0.55, 0.5, 0.45, 0.4, 0.35, 0.3, 
0.25, 0.2, 0.15, 0.1, 0.05, -0.05) percent of the peak surface brightness of 
96.2 mJy~beam$^{-1}$ at 3.8 cm.
Lines show the magnetic field vectors. Ticks in (a)-(c) are every 2\arcsec,
N is at top, and each panel is 20\arcsec\ on a side.
\label{fig:radio}}
\end{figure}

\begin{figure}
\includegraphics[scale=0.84]{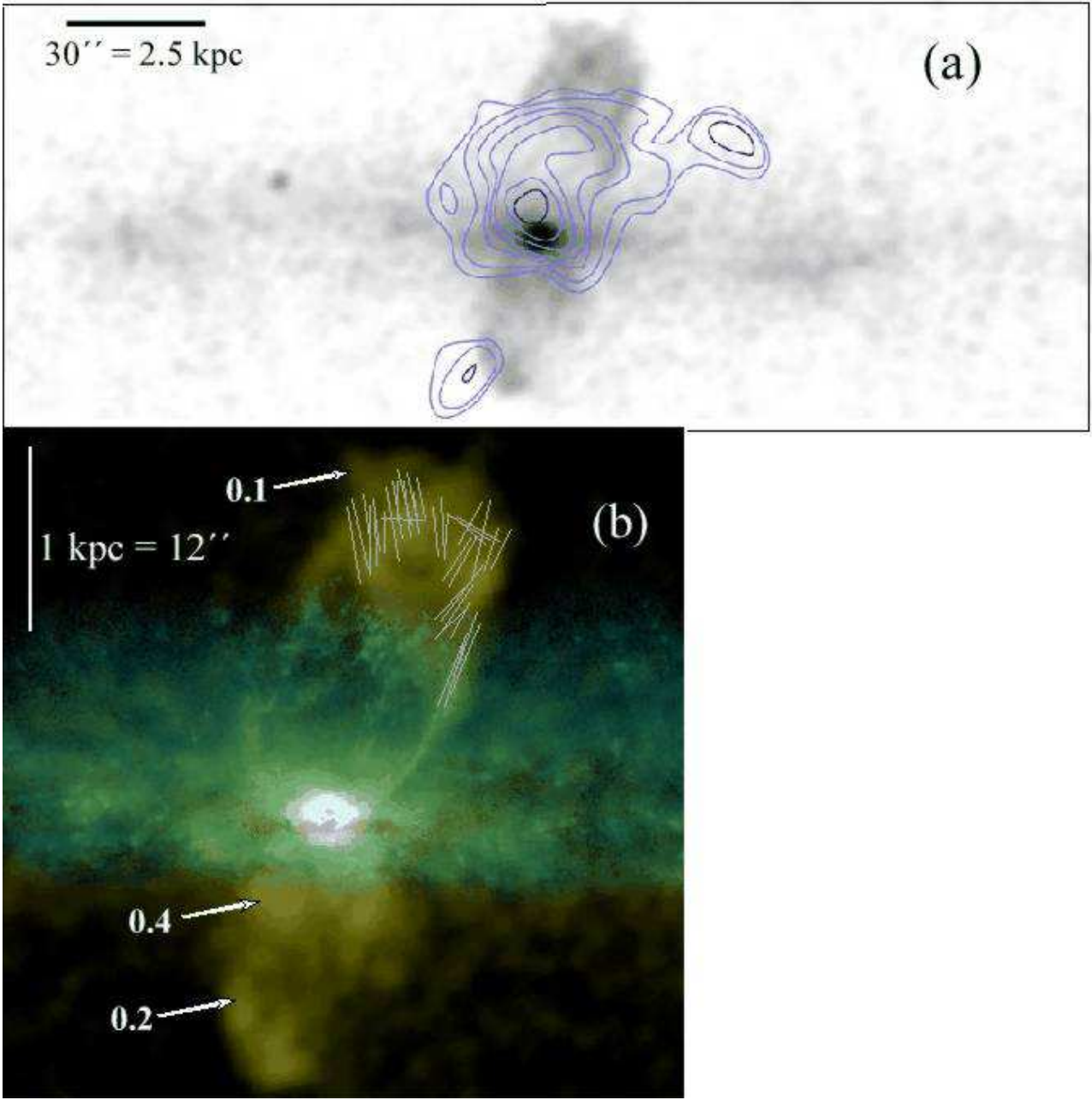}
\caption{
(a) The 6 cm VLA image with \emph{ROSAT} HRI contours across the central part. 
The radio beam is \protect$ 1\farcs 59\times 1\farcs 65.\protect $ 
The region shown is 4\arcmin\ wide, and is oriented like
Fig. \ref{fig:overall}. Same contours are shown as in Fig. \ref{fig:bubble}b.
(b) Line
emission (green) is superimposed on the radio continuum (yellow) image in (a).
Representative areas are labeled in units of \protect$ 3.7\times 10^{-4}\protect $
Jy~arcsec\protect$ ^{-2}\protect $. Magnetic field vectors, obtained by
fitting P.A. vs. \protect$ \lambda ^{2}\protect $, are drawn on the radio
structure. The polarization of the bright knot at top is uncertain due to 2\protect$ \pi \protect $
ambiguities.  \label{fig:radiopol}}
\end{figure}

Radio emission traces the outflow on all observed scales, and many images have
been made following the discovery of minor-axis radio lobes by \citet{Du83}.
Using VLA images at 1.4 and 4.9 GHz, they constrained
the rotation measure to the E bubble,
$50 < R_m < 300$ rad m$^{-2}$. But polarization studies are tough at 1.4 GHz
because the 30\% polarization seen at 4.9 GHz has been reduced to $<$3\%
by Faraday rotation.
% AC277 and AC298 programs
From the many datasets of NGC 3079 in the VLA archive, we therefore
extracted unpublished datasets made by Dr.\ S.\ Caganoff and collaborators
which are centered at 8.1 and 4.4 GHz
(Fig.\ \ref{fig:radio} and Fig.\ \ref{fig:radiopol}a).
Details of the observations are given in Table 2.

Standard procedures under AIPS reduced the data at four frequencies. We then
fit $\chi(\nu)= \chi(0) + \lambda^2 R_m$ to each
pixel to map the transverse electric field ($E_\perp$) $\chi$(0) and the $R_m$.
Uniform weights produced a beam of 1.65\arcsec 
$\times$ 1.59\arcsec\ (HPBW) along PA = 13.3\arcdeg.
Figs.\ \ref{fig:radio} and \ref{fig:radiopol}
are insensitive to features $>$30\arcsec.
Fig.\ \ref{fig:radiopol}b shows that the 3.8 cm 
radio emission is $<3.4\times 10^{-5} $ Jansky~arcsec$ ^{-2} $ inside the
superbubble.
\placefigure{fig:radio}
\placefigure{fig:radiopol}

\section{\label{sec:bubbledata}EMPIRICAL RESULTS}

\subsection{\label{sec:dust}Dust Plumes and the Pattern of Scattered Light in the Galaxy
Bulge}

Much of the {}``peanut bulge{}'' discussed in Paper IV may arise from scattered
nuclear or inner-disk starlight. \cite{Is98} argue from their near-IR broad-band
images that 20--30\% of the bulge light must be scattered. The pattern we see
in the I-band image (Fig.\ \ref{fig:overall}a) is bilobal and straddles
the superbubble. The outflow appears to have depleted the central
cylinder of dust, and indeed, several dust plumes arch out from
the bottom half of the superbubble to fall back onto the galaxy disk. The increased
dust load in the presumed toroidal lobe enhances scattered starlight. Scattering
in the foreground halo also seems to {}``soften{}'' the appearance of dust filaments
near the nucleus and in the background disk to produce the striking impression
of ``depth{}'' in e.g.\ Fig.\ \ref{fig:bubble}a.

We quantified dust extinction by comparing the \emph{I}-band surface
brightness measured toward a dust feature $ S_{dc,I} $
with its local background $ S_{bg,I} $ \citep[e.g.][]{Ho97}.
Foreground starlight reduces absorption by
$ A_{I}=-2.5\log (S_{dc,I}/S_{bg,I}) $.
Defining $ x $ as the fraction of starlight emitted in front of the
feature, 
$ S_{dc,I}/S_{bg,I}=x+(1-x)e^{-\tau _{I}} $ where $ \tau _{I} $
approximates the maximum extinction optical depth through the dust. We
ignored light forward scattered along our line of sight, and
also assumed that dust has small depth compared to stars so it acts
as an external screen. 
We averaged 5-pixel wide intensity cuts through a prominent
dust arc in Fig.\ \ref{fig:bubble}a at the edge of the superbubble.

A single color cannot constrain tightly both $ x $ and $ \tau _{I}$.
By bounding $ x\,\ (0.4<x<0.8) $, we found that $ 0.27<A_{I}<1.6 $ mag. 
($ 0.48<A_{V}<2.8 $ mag, assuming $ R=3 $ for the ratio of absolute
to relative extinction).
Multiplying $ N_{H}=1.7\times 10^{21}A_{V} $
cm$ ^{-2} $ by $ 1.37 $ to convert H to total mass (He + typical metal
fraction), and treating the dust arc as a cylinder of radius $ 0\farcs 3 $
and length 3\arcsec, we found that the \emph{total} mass is $ (1-4.7)\times 10^{5}M_{\bigodot }, $
larger if dust was destroyed during cloud acceleration. Despite
this high mass, the dusty feature shows no line emission, underscoring the need
to map the outflow in many ISM phases.

\subsection{Dereddening}

The \emph{J-K} color map of \cite{Is98} shows enhanced reddening W of the nucleus.
This is also apparent in Fig.\ \ref{fig:overall}b as an over 
subtraction of the I-band image from the F658N exposure, which forms
the dark zone at the
base of the superbubble. Fig.\ \ref{fig:bubble}a shows that the ionized filaments
of the superbubble originate from
an elliptical region $\approx7\arcsec \times 3\arcsec$ in extent,
with patchy obscuration.
On this scale there is a ring between 1\farcs25
\cite[from CO]{So92}
and 3\farcs5 \citep[from molecular H and radio continuum]{Me98} radii that
is rotating at $ \approx 330 $ km~s$ ^{-1} $ \citep{Is98}. 
\cite{Is98} find $ H_{2}(v=1\rightarrow 0)S(1) $ emission of $ \approx 10^{7} $
L$ _{\bigodot } $ to a height of $\approx100$ pc and out to 300 pc near the
galaxy disk plane.

In Paper II we used long-slit spectra of the H$ \alpha  $ and
H$ \beta  $ lines (corrected for Balmer absorption) to determine an average
reddening optical depth of $ \tau  $$ (H\alpha )\approx 5 $ at the nucleus,
$ \approx 1 $ in the bottom third of the superbubble, and zero beyond;
the value on the nucleus agrees with that found by
\cite{Is98} across the inner $ 6\arcsec \times 2 $\arcsec.
Accordingly,
to deredden the filaments, we bracketed the extinction gradients with height
by using two linear trends such that $ A_{V}=2.75 $ at the base of the bubble,
declining to $ A_{V}=1.75 $ at heights between 3 and 4\arcsec, and to no
reddening by 8\arcsec\ and 10.5\arcsec\ height, respectively. 
The superbubble is projected on diffuse, ionized emission which is a mixture
of true vertical structure and the outskirts of the background galaxy disk.
As we summed the flux of each filament, we therefore subtracted the average
vertical gradient outside the superbubble. 

\begin{figure}
\includegraphics[scale=0.83]{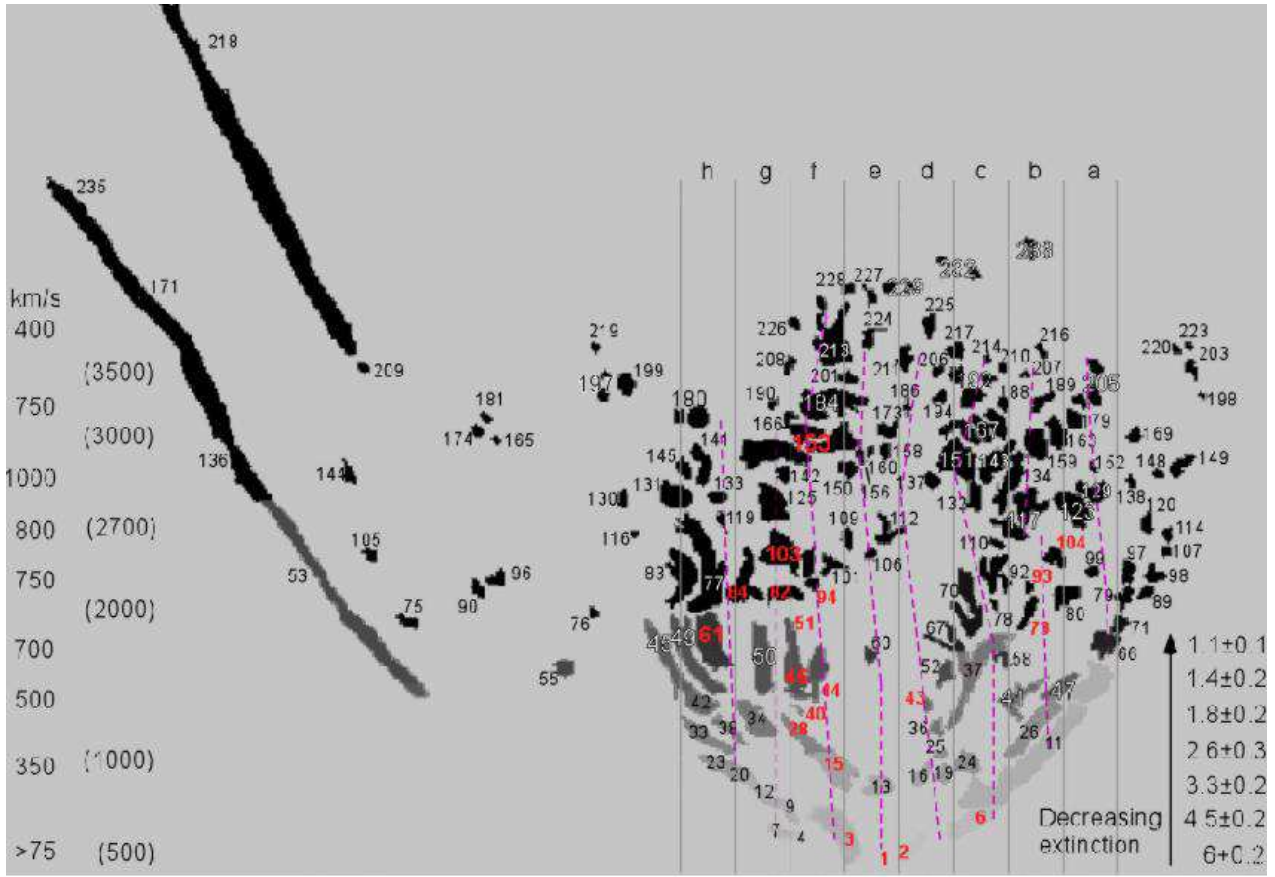}
\caption{
Filaments in the \emph{HST} line image are numbered for reference in the list
of their physical properties, Table 3.
The orientation is that of Fig.\ \ref{fig:bubvels}, 
with P.A.\ 80\arcdeg\ at top and ticks every 0\farcs5.
The vertical lines show the
effective {}``slit width{}'' and sampling of our imaging Fabry-Perot
spectrophotometry (when binned as in Fig.\ \ref{fig:bubvels}). Within each  ``slit{}'' the
dashed line shows the variation of radial velocity in the range 0 (right hand
side) to -1000 (left) km~s\protect$ ^{-1}\protect $ relative to galaxy systemic.
These trends are used to assign first an average radial velocity to each filament
(see text), then --- assuming cylindrical symmetry --- the \emph{minimum} 
space velocity
shown at the left-hand side of this figure. For comparison, the velocities used
in Paper II are shown in parentheses. Filaments with negative (approaching)
radial velocities are labeled in black or white, receding velocities are shown
in red. We assume that extinction varies vertically across the 
superbubble, and show at bottom right the multiplicative
factor used to deredden the H\protect$ \alpha \protect $ fluxes at
each height for the range of reddenings
consistent with the observed Balmer emission-line decrements (Paper II).
\label{fig:label}}
\end{figure}

Isolated in this
fashion, the superbubble H$ \alpha  $ flux sums to $ \approx 10^{-13} $
ergs~s$ ^{-1} $~cm$ ^{-2} $ with compact filaments emitting slightly
more than half, the jet (\S\ref{sec:jetgas}) 10\%, and diffuse emission
1/3, of the total. 
We found an H$ \alpha  $ luminosity of the filaments
of $ (3.5-3.9)\times 10^{39} $ ergs~s$ ^{-1}$ (depending on the
reddening), comparable to 
what we derived from the ground-based spectra (Papers I \& II). 
\placefigure{fig:label}
\placefigure{fig:halum}

\subsection{\label{sec:ke}Constraints on Ionized Densities and Masses}

\begin{figure}
\includegraphics[scale=0.55]{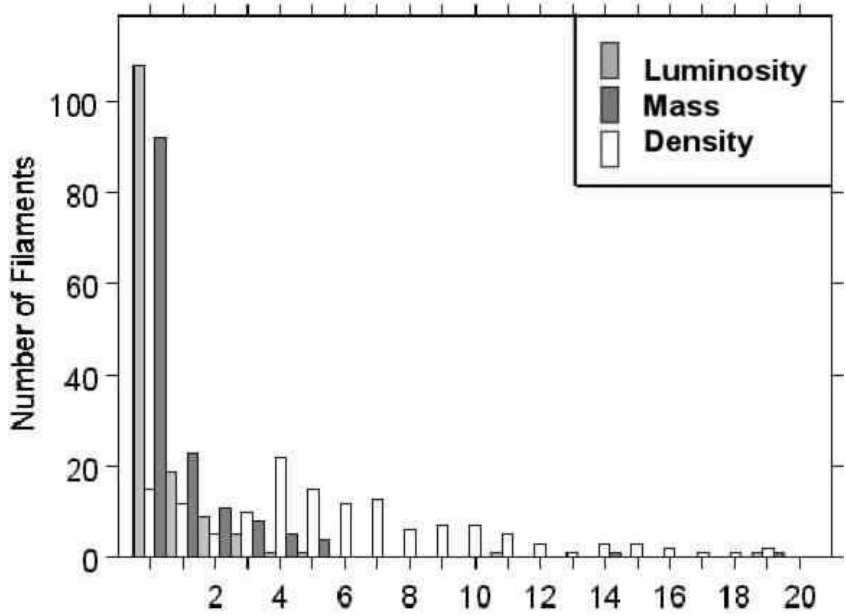}
\caption{
Distributions of the derived properties of the
ionized filaments in the superbubble. H\protect$ \alpha \protect $
luminosity is in units of 10\protect$ ^{38}\protect $ ergs~s\protect$ ^{-1}\protect $, and
assumes the smaller reddening correction (see text). Electron density  is in
units of \protect$ f^{-1/2}\protect $ cm\protect$ ^{-3}\protect $.
Ionized mass  is in units of \protect$ 10^{4}\protect $ \protect$ \sqrt{f}M_{\bigodot }\protect $, and
assumes case-B recombination conditions at 10$^4$ K.
\label{fig:halum}}
\end{figure}

Fig.\ \ref{fig:halum} shows the H$ \alpha  $ luminosity function of filaments
labeled in Fig.\ \ref{fig:label},
assuming the smaller reddening discussed above.
We estimated a \emph{minimum} volume of each filament
by assuming a cylinder with depth equal to its thinnest dimension (measured
at the contour of 50\% peak emission) on the sky; many filaments are barely
resolved in one dimension --- $ \approx  $0\farcs25 FWHM thick after accounting
for instrument blur --- and
most at the top of the superbubble are unresolved.
Assuming Case-B recombination conditions at 10$ ^{4} $ K and $ n_{p}\approx 0.9n_{e}, $
the electron density is $ n_{e}=1.8\times 10^{12}\sqrt{L_{H\alpha }/Vf} $
cm$ ^{-3} $ for gaseous filling factor $ f $ and emitting volume $ V $.
The flux-averaged, ionized density of the superbubble filaments is $ n_{e}=5.7f^{-1/2} $
cm$ ^{-3} $; Fig.\ \ref{fig:halum} shows the population distribution and
the distribution of ionized mass which sums to
$ 1.4\times 10^{6}\sqrt{f} $ M$ _{\bigodot } $. We used Monte Carlo techniques
to quantify the uncertainties of these derived parameters. We
assumed that the H$ \alpha  $ luminosity is known to $ \pm 25\% $ (half
of this uncertainty comes from the reddening ambiguity, half from the distance
to NGC 3079) and the volume to $ \pm 50\% $ (which encompasses the range
of filament widths that are resolved in the H$ \alpha  $ image). 
Table \ref{tab:parms}
lists mean and extreme values of the density and mass of each filament under
these assumptions, and whether the filament is resolved.
\placetable{tab:parms}

In Paper II we found no evidence from the red {[}\ion{S}{2}{]} doublet flux ratios
that the ionized gas exceeds the low-density limit $ n_{e}\ga 100 $ cm$ ^{-3} $
across the superbubble. This and our new \emph{HST} results
now constrain the gaseous filling factor $ f $ to $ >0.3\% $ of the gas volume.
A typical filament therefore has ionized mass $ \ga 300M_{\bigodot } $; filaments
near the top of the superbubble are smaller and fainter, and we find
have ionized mass $ \ga 60M_{\bigodot }. $ If filaments are fragments of
molecular clouds lofted from the galaxy disk
\citep[see e.g. the models of][]{Sc86,Su94}, 
then this constrains only the mass of their outer ionized sheath.

\subsection{Superbubble Filaments}

Fig.\ \ref{fig:bubble}
shows that four prominent bundles of ionized filaments emerge from the CO ring,
and appear to be
spaced $ \approx  $90\arcdeg\ apart in azimuth.
They are close together in (perhaps intertwined?) bundles
until they reach a
height of 550 pc, at which point each bundle separates and the strands,
otherwise unchanged in appearance, make sharp angles on the sky
while still trending upward.
Fig.\ \ref{fig:radiopol}b shows that at greater heights strands are replaced
by a spray of loops and arches that appear to {}``drip{}'' toward the galaxy
disk; the apparent connection between knots
and arches may be spurious because of limited spatial resolution. 
Fig.\ \ref{fig:bubble}b shows that the soft X-ray emission peaks along the drips.
Beyond, isolated small clouds are evident $ > $1.5 kpc above the galaxy disk.
Beyond the clouds, radio emission continues outward
as a {}``cap{}'' to 28\arcsec\ (2.3 kpc) radius.
At still larger radii, soft X-ray emission
fades into the \emph{ROSAT} background (PTV).

\subsection{\label{sec:radiostuff}Constraints on the Magnetic Field}

Fig.\ \ref{fig:radio}c shows that
the E radio lobe has a well-ordered magnetic field,
indicated both by its high polarization ($20-30$\% at 6 and 3.8 cm
wavelengths) and by the trend in rotation measure
$R_m$.  While $R_m$ is nearly constant along the polar
axis, perpendicular it changes sign from $\sim~+$50 
rad m$^{-2}$ on the inside to $\sim~-50$ rad $^{-2}$ toward the outer edge. 
The line-of-sight component of the magnetic field ($B_\parallel$) reverses
dramatically whereas the transverse field $B_\perp$ 
($= E_\perp + \pi/2$) aligns with the outflow, particularly 
along the N edge.

The E lobe appears as a loop or limb-brightened
shell; similar radio structures are seen in other inclined galaxies 
\citep[e.g. NGC 2992,][]{We87}.
Whether it forms part of a shell, closed or open
loop or bubble, anchored or otherwise to the central disk is
an important clue to its origin. Unfortunately, Fig.\ \ref{fig:radio}
shows that we have information only for the sides of the radio lobe.
The connection with the disk and the cap of the radio lobe (Fig.\ \ref{fig:radio}a)
is too faint to derive polarized intensities. However, the 
$R_m$ inversion $-$ inside to outside $-$ is compelling and, to our knowledge,
has never before been observed.

The form of $R_m$ excludes a field centered on
the lobe because the projected field would integrate to zero.
Our fits to $R_m$ follow a $\lambda^2$ dependence closely, all but ruling out
a Faraday-rotating medium within the synchrotron source \citep{So98}.
Because polarizations are sometimes close to the maximum possible 
value \citep[75\%,][]{Bu66}, little depolarization can have occurred.
Indeed, the slow spatial/spectral behaviour of $R_m$ compared 
to the beam rules out beamwidth/bandwidth effects.  Thus, the Faraday screen 
lies predominantly in front of the synchrotron source and is partly resolved.
The mean $R_m \approx 50$ rad m$^{-2}$ is easily
generated and constrains the density of the screen.
\citet{Du83} derive a mean field of 20 $\mu$G for the E
lobe. We derive a lower limit on $B_\parallel$ by equating $\vert R_m \vert$ 
$\approx$ 50 rad m$^{-2}$ with the limit on H$\alpha$ emission measure in the
lobe ($E_m$ $\le 1$ cm$^{-6}$ pc). If the depth of the Faraday screen is 
$L$ $\sim$ 10~pc, because $R_m = 0.81 B_\parallel n_e L$, we derive 
$B_\parallel \ga 15$ $\mu$G. The required gas density is $n_e \la 0.3$ cm$^{-3}$.

At the wind speed postulated in \S4.2.1, it is unlikely that
magnetic energy density dominates the thermo-kinetic energy 
of the flow. The wind may simply carry along the magnetic field
\citep{Kl88}.  Unless the compressed field is much higher than implied by
minimum energy arguments, the above numbers are fairly restrictive and 
suggest that the radio lobe beyond the superbubble should emit detectable
H$\alpha$ emission
at a flux of 10$^{-18}$ ergs cm$^{-2}$ s$^{-1}$ arcsec$^{-2}$, 
$\approx20\%$ of our current $2\sigma$ threshold (Paper II).

In light of the optical data, we suspect that the polarized emission arises 
from a limb-brightened incomplete shell (partial bubble)
that is tilted slightly to the line of sight.
The radio bubble is expanding into a magnetized, hot (PTV), low-density 
halo.  The radio 
continuum is presumably limb-brightened where the field is highly compressed. 
In \S\ref{sec:radiodiscuss}
we provide a qualitative model to explain the $R_m$ inversion.

\subsection{Optical Emission from the Counterbubble}

In Fig.\ 11 of Paper II we traced line emission from high-velocity gas
between projected radii of 10\arcsec\ and 15\arcsec\ (panels c \& d) W of the
nucleus.
Here Fig.\ \ref{fig:overall}b shows
line-emitting regions with [\ion{N}{2}]$\lambda$653/H$\alpha$ flux ratios
like those in the E superbubble. Spectra of IR
recombination-line flux ratios
would constrain the reddening, hence energetics of the
ionized counterbubble.

\subsection{Morphology of Disk H II Regions}

Properties of the star-forming, large-scale galaxy disk
will be considered in a later paper.  However, energetic,
non-planar motions are of immediate relevance.

A notable feature of the H~II regions in NGC 3079 is their
tendency to {}``blister open{}'' at their tops, see especially the outer ring
in the right-hand half of Fig.\ \ref{fig:overall}b. H~I arc {}``C{}'' of
\cite{Ir90} lies above this region. As discussed in \S4.4 of Paper III, such
structures suggest vertical motion of gas up {}``galactic chimneys{}'' which
also allow photons to escape from the disk to energize the
prominent diffuse ionized medium.

\begin{figure}
\includegraphics[scale=0.44]{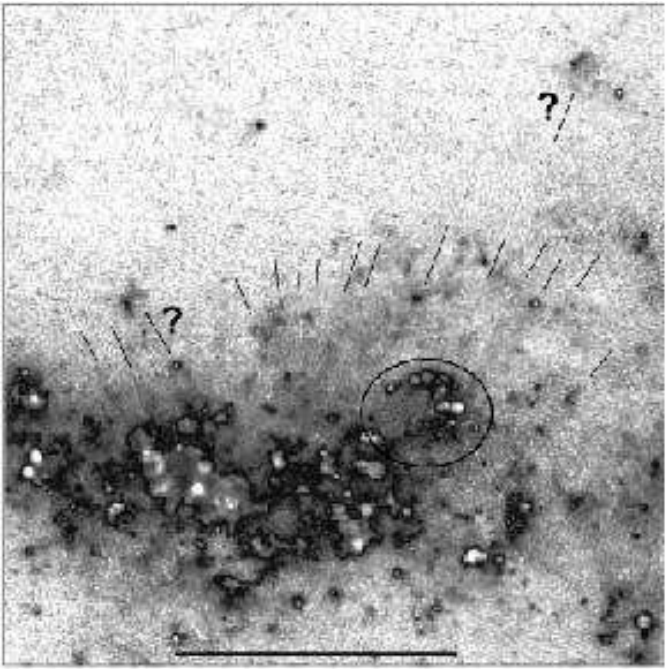}
\caption{
Linear filaments marked with lines
emanate from a major star-forming complex in this line-emission image.
The area shown is the 40\arcsec\ wide boxed region
near the left-hand edge of Fig. \ref{fig:overall}b, and lies beyond the maximum
radius of the stellar bar. The line at bottom spans 15\arcsec\=1.25 kpc.
Two larger, apparently edge-brightened nebulae that
may be associated with this outflow are marked with {}``?{}''. A structure
resembling a star-forming ionization front is circled. \label{fig:blast}}
\end{figure}

A striking active region is marked
near the left-hand side of Fig.\ \ref{fig:overall}b, where
Fig.\ \ref{fig:overall}a shows a high density of resolved stars
and dust filaments.
This complex
lies well beyond the maximum radius of the stellar bar, is a local flux peak
in the 245 GHz continuum (8 mJy over 11\arcsec\ diameter), lies below H~I
arc {}``D{}'' of \cite{Ir90} which requires formation energies $ (0.5-2)\times 10^{54} $
ergs,
and has X-ray luminosity $ \approx 5.4\times 10^{38} $ ergs~s$ ^{-1} $
in the \emph{ROSAT} HRI band (source H14 of PTV).  Fig.\ \ref{fig:blast}
shows that over a diameter of 15\arcsec, roughly a
dozen linear filaments protrude for 0.6 kpc. They 
have broad line widths ($ \approx  $200 km~s$ ^{-1} $ corrected
for instrumental resolution), but mean velocities
equal to that of the rotating galaxy (Paper IV for the ionized gas, and \cite{Ir91} for
the H~I) at their projected radius. 
Fig.\ \ref{fig:blast} also shows 
unresolved knots of line emission at the ends of many of these filaments.
The three brightest knots have H$ \alpha  $
luminosities of $ (1.5\pm 0.1)\times 10^{37} $ ergs~s$ ^{-1} $ assuming
no reddening. Two features labeled {}``?{}'' in the figure have velocity
FWHM $ \approx  $200 km~s$ ^{-1} $ in our Fabry-Perot spectra (Fig.\
4c of Paper III), and sharp edges on the side directed away from the source
of the linear filaments. 
These structures resemble superficially Galactic Herbig-Haro
objects, but are far more energetic.
\placefigure{fig:blast}

\subsection{\label{sec:velfield}Velocity Field of Ionized Filaments in the
Superbubble}

\subsubsection{Previous Results}

As mentioned in \S3.2,
emission line profiles in Fig.\ \ref{fig:bubvels} have a component
from gas in the background galaxy that is separable from energetic
gas in the superbubble by its modest deviation from the galaxy systemic velocity
\footnote{
As discussed in \S3.2.1 of Paper IV, the systemic velocity of NGC 3079
depends on the waveband of observation.
Values 1150$\pm$25 km~s$^{-1}$ result
from the combined effects of a disk bar and possible warp, and
differences in optical depth and orientation to various galaxy components.
We use $v_{sys}=1150$ km~s$^{-1}$.}
and by the small values of its {[}\ion{N}{2}{]}$ \lambda  $6583/H$ \alpha  $
flux ratios (which are typical of H II regions, see Fig.\ 3 of Paper II).
The spectral profile from a 0\farcs57 high$ \times  $2\farcs3 wide box often
spans several hundred km~s$ ^{-1} $, arising from {}``turbulent{}''
motions within a filament or gradients that are unresolved at ground resolution.
An example is at the bottom left-hand corner of the superbubble,
where filaments 45, 49, \& 61 have opposite velocities ($ \pm 150 $ km~s$ ^{-1} $)
despite being adjacent on the sky.

In Paper II we showed that the superbubble velocity field is a partial {}``Doppler
ellipsoid,{}'' with the centroids of emission-line components spanning
radial velocities -1050 to +550 km~s$ ^{-1} $ relative to $v_{sys}$,
and having maximum velocity splitting 1250 km~s$ ^{-1} $ near the
mid-axis of the superbubble. 
Velocities do not return
to small values at the top of the superbubble as they do in spatially resolved
spectra of planetary nebulae; instead, line emission fades away as velocities
reduce modestly from extreme values. In Paper II we showed that this pattern
is quite well reproduced by a nuclear outflow wherein filament space velocities
accelerate to 3700 km~s$ ^{-1} $ at $ r=1 $ kpc as $ v(r)\propto r^{\alpha } $,
with $ \alpha \sim $2.5, while motion vectors swing toward the sky plane
as the bubble cap is approached. The main deficiency is in the N half of the
superbubble, where the model
predicts redshifts from v$_{sys}$ 
of 550 km~s$^{-1}\ vs.$ 200 km~s$^{-1}$ observed for the presumed backside gas.

\begin{figure}
\includegraphics[scale=0.83]{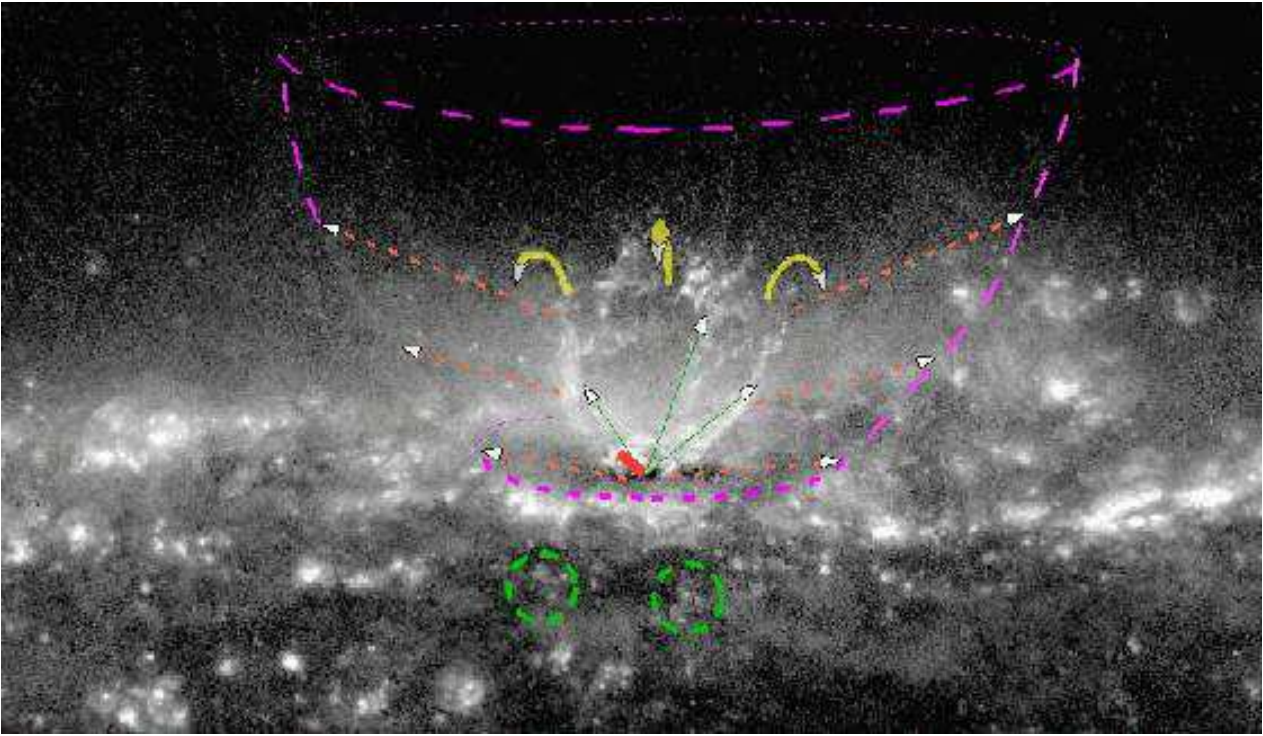}
\caption{
Inferred velocity fields associated with the nuclear
outflow. The orange dashed lines point to highly inclined ionized filaments
that may form the stagnation boundaries of the nuclear wind as it blows
through the rarefied component of the galaxy ISM. The green lines point
to the dense component of the disk ISM which has been elevated to form the superbubble.
The bubble has ruptured into vortices at its top (schematic motions are shown in
yellow). The red line at its base shows the filament stream that aligns with
the VLBI-scale jet (see Fig. \ref{fig:closeup}). Circled at bottom in 
dashed green are filaments whose emission-line spectra show high velocities
and non-H-II region excitation (Paper II) perhaps
associated with the optically
extinguished radio counterbubble.  \label{fig:interpretation}}
\end{figure}

\placefigure{fig:interpretation}

\subsubsection{\label{sec:velmodel}Refinement Motivated by Present Data}

The \emph{HST} line image shows that the filaments are organized into
four well-defined vertical bundles, each fortuitously spanning
the 1\arcsec\ resolution
of our ground-based spectra. Although we cannot assign an unambiguous value
to a filament, we \emph{can} bound velocities in each bundle
with the radial velocities mapped in Fig.\ 4 of Paper II.
We find that most filaments have blueshifted motions relative to $v_{sys}$,
with largest radial velocities in the spray of 
arched filaments near the top of the bubble.
Anticipating the 
dynamics to be discussed in \S\ref{sec:discuss}, it is plausible that
these are {}``mushrooming{}'' out of the bubble with a substantial motion
along our sightline. So we refined the field used in Paper
II, now assuming expansion mostly parallel
to the disk plane at set height in the
bubble (Fig.\ \ref{fig:interpretation}), with space velocity comparable to  
the observed velocity of the blue wing along the mid-axis of the bubble.
At left in Fig.\ \ref{fig:label} we show 
the adopted space velocity at each height.
The expansion velocity that we infer at the \emph{base} of the bubble
is consistent with the values
derived there by \cite{Ha95} from $ H_{2} $ transitions and
mapped by \cite{Is98} (see also Fig.\ \ref{fig:closeup}).
Together, this model and the nucleus-centered pattern used in Paper
II should bracket the true motions. 
\placefigure{fig:closeup}

\subsection{Bulk Energies and Momenta of the Ionized Filaments}

Table \ref{tab:parms} lists maximum and minimum space velocities of each filament as derived from
the projections of Paper II and \S\ref{sec:velmodel}, respectively.
Total KE
and momentum of ionized gas in the superbubble sum to
$ (2.9-0.26)\times 10^{55}\sqrt{f} $
ergs and $ (3.4-0.74)\times 10^{47}~\sqrt{f}
$ dyne~s, respectively; values for each
filament are also shown in Table \ref{tab:parms}. 
We retain our estimate from Paper
II of {}``turbulent{}'' energy --- $ 3.1\times 10^{54}/n_{e} $ ergs for
the blue-wing component. This is now an upper limit, because we now resolve
some of the line broadening into discrete filaments each with
a bulk KE.

\subsection{\label{sec:jetgas}Jet Emission Near the Nucleus}

\begin{figure}
\includegraphics[scale=0.8]{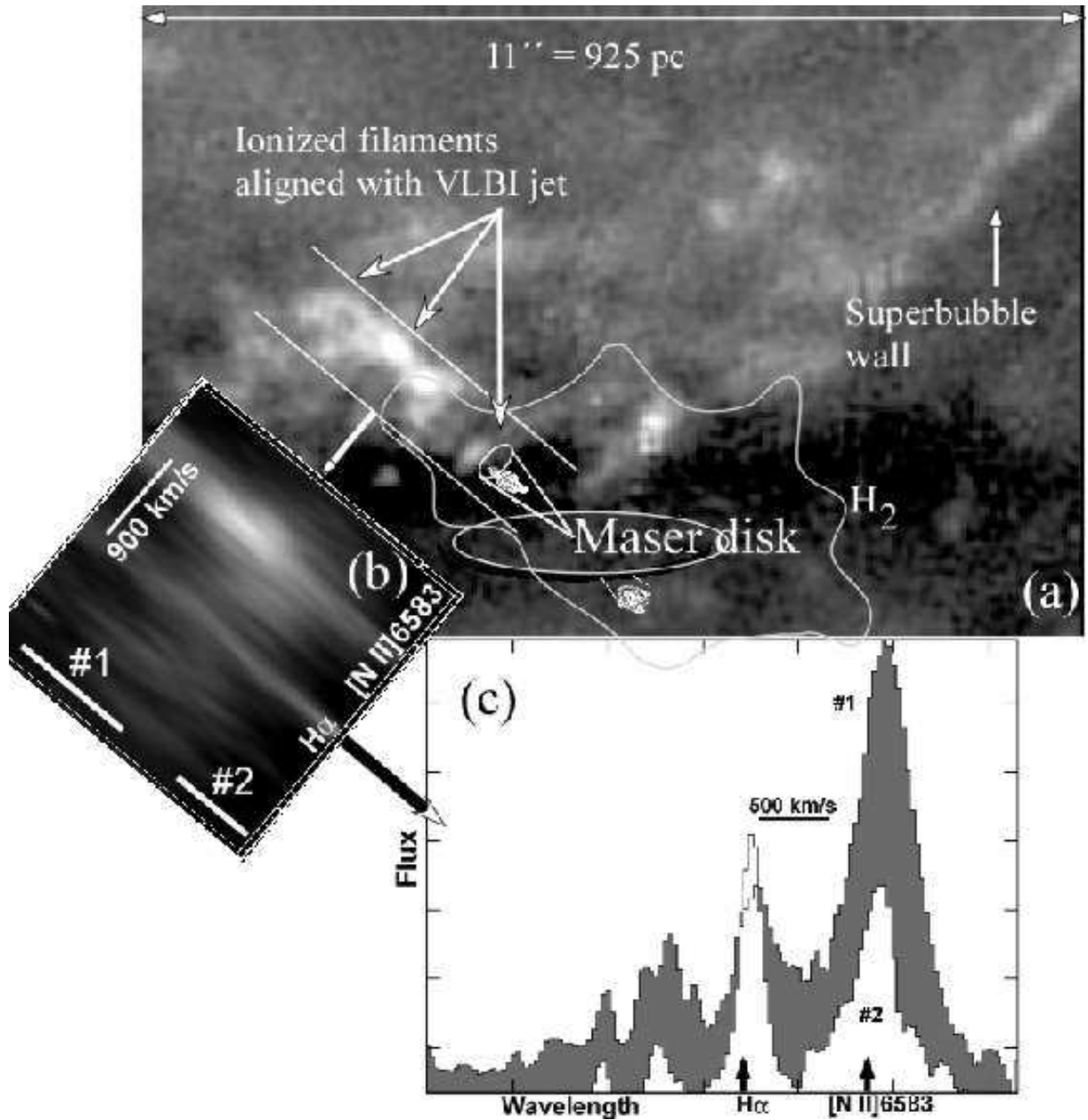}
\caption{
(a) The base of the superbubble in H\protect$ \alpha \protect $
emission near the nucleus, shown with the
same orientation as Fig.\ \ref{fig:bubvels}. The
image was smoothed with a Gaussian kernel of 0\farcs18 FWHM. The brightest filament stream
trends toward the S (left) boundary of the superbubble, and coincides in projection
with the putative
VLBI-scale jet, which is shown magnified \protect$ 250\protect $
times using the image of \cite{Tr98} 
(not all radio-jet knots are shown). The bottom of the image is obscured
by a dusty molecular ring.
(b) Shown is the space-velocity diagram of the {}``long-slit{}'' spectrum 
extracted
from our Fabry-Perot datacube along the jet in a 1\farcs7-wide box.
The two most prominent spectral lines are labeled, while the white lines at
lower left show the regions over which the jet emission-line profiles were extracted
for display in panel (c). (c) Emission-line profiles averaged from the regions
shown in (b). Each spectral bin is 0.57 {\AA} wide (\protect$ \approx \protect $33
km~s\protect$ ^{-1})$.
Arrowheads indicate $v_\textrm{sys}$ for the line species,
with the wavelength scale increasing to the right.\label{fig:closeup}}
\end{figure}

Fig.\ \ref{fig:closeup} shows that the brightest stream of
ionized filaments projects
near the S boundary of the superbubble base, and coincides with the axis
of the putative VLBI-scale jet mapped at 8 GHz by \cite{Tr98} and
\cite{Sa00}. In the
magnified inserts in Fig.\ \ref{fig:overall}
this feature has H$ \alpha  $ surface brightness
$ \approx 5\times 10^{-15} $ ergs~s$ ^{-1} $~cm$ ^{-2} $~arcsec$ ^{-2} $
coincident with
a narrow plume of I=16 mag arcsec$ ^{-2} $ continuum light that points
to the nucleus.
Using the \emph{HST} line-only image Fig.\ \ref{fig:overall}b
to define a mask of the jet emission,
we summed spectra from our ground-based Fabry-Perot datacube to map
velocities (see Fig.\ \ref{fig:closeup}b), then averaged
spectra at two intervals along the jet (Fig.\ \ref{fig:closeup}c); 
the mean velocity of the two summed spectra is redshifted by 125
km~s$^{-1}$ relative to v$_{sys}$.
Elsewhere the corrected FWHM is 200 km~s$ ^{-1}$ and
[\ion{N}{2}]/H$\alpha \approx1.2$. In the two summed spectra we subtracted 
this component to isolate a broader 
jet-specific feature whose H$\alpha$ flux we then deblended from
the adjacent
{[}\ion{N}{2}{]}$ \lambda \lambda  6548,6563$ doublet.  This
component has centroid \emph{blueshifted} 125 km~s$^{-1}$ relative to 
v$_{sys}$.  Fit results are reported in Table \ref{tab:jetparms};
the KE is a lower limit assuming
that the broader feature arises entirely from expansion
around the jet that projects all motion onto our line of sight.
Actually, gas kinematics constrains jet orientation and 
space velocity poorly because gas is ionized at
radiative shocks within the thermally unstable cocoon/ISM interface.
\placetable{tab:jetparms}

\section{DISCUSSION}

\subsection{Nuclear Jet}

\cite{Ir88}
suggested that the superbubble could have been blown by a precessing
VLBI-scale jet; \cite{Ba95} elaborated on this scenario.
\cite{Ir92} interpreted extensions of CO(J=1-0)-emitting gas into the
base of the superbubble as resulting from a jet/ISM cloud
interaction. Using pressure confinement arguments and the minimum momentum flow,
they estimated that the putative jet has velocity $ >10 ^{4} $ km~s$ ^{-1} $
and power $ >4\times 10^{40} $ ergs~s$ ^{-1}, $ respectively. Over the 1
Myr dynamical age of the superbubble (Paper II)
the KE exceeds $ 10^{54} $ ergs, which suffices to power the CO outflow.

\cite{Ir88} resolved the nuclear emission into a flat-spectrum radio
core with parsec-scale collimation along P.A. 120\arcdeg. \cite{Tr98} further
resolved this structure into four clumps with radio spectra
inverted between 5 and 8 GHz, and steep between 8 and 22 GHz; in this sense
the nuclear sources resemble a
low-power analog of powerful radio galaxies. \cite{Sa00} found that the
spectral indices and fluxes of several of these knots changed considerably
in 22 months and that knot motions were consistent with 
expansion from the nucleus at an apparent velocity of $\sim0.16$c.
The four clumps align precisely
(see \S\ref{sec:jetgas} and Fig.\ \ref{fig:closeup})
with the brightest stream of filaments at the base of the superbubble. 
Three
results suggest that the elongated, ionized feature is energetic (not
ambient gas photoionized by the AGN): 

\begin{enumerate}
\item Fig.\ \ref{fig:bubble}b shows that X-rays
peak off-nucleus. \emph{ROSAT} was pointed to $ \approx \pm 2 $\arcsec\ accuracy
on this galaxy (PTV), about the displacement from the nucleus in
the Figure. \cite{Pa97} modelled the \emph{ASCA+ROSAT} PSPC spectrum to show that
the nucleus is quite obscured
($(1-2.3)\times 10^{21}$ cm$^{-2}$), so the jet
may be generating these X-rays.
\item Velocities of both ionized (\S\ref{sec:jetgas}) and molecular gas \citep{Is98}
peak in this extranuclear region.
Velocities of order $V_{s,400} = 400$ km~s$^{-1}$ will produce shock emission
$ L(H\alpha )_{shock}=1.5\times 10^{34}n_{0}AV^{3}_{s,400} $
ergs~s$ ^{-1} $ (\emph{A} is the shock surface area in pc$ ^{2} $
in an ambient ISM of density $ n_{0} $ cm$ ^{-3} $).  The luminosity of
filament \#3 in Table 3 arises from pure shock ionization
if $n_0\sim4$ cm$^{-3}$.
\item
Here optical emission-line profiles
are broad ($\sim440$ km~s$^{-1}$ FWHM,
\S4.1), as we would expect from gas shock-heated as a jet
impacts the inner boundary of the molecular cavity in the
disk. Indeed, \cite{Is98} see bright H$ _{2} $ emission and hot
dust $ T\approx 900\pm 100 $ K nearby.
\end{enumerate}

\cite{Ro20} showed by numerical simulation that
a jet propagating through the ISM in a spiral galaxy
radiates on average $ \approx 0.1-0.5\% $ of its KE as ionizing photons at each
interaction site, with ionizing luminosity $\approx15L_{H\beta}$
over a wide range of jet velocities and ISM densities.
From Table 3 the ionizing luminosity
$1.7\times 10^{39} $ ergs~s$ ^{-1} $ can therefore inject the KE
$\approx 5\times10^{52}/0.005=10^{55}$ ergs of the superbubble
over its dynamical age if the jet has operated with high enough duty cycle.

We can constrain the age of the radio bubble independent
of the optical outflow.  If we assume the bubble was inflated 
by the jet but then lifted by buoyancy, the buoyancy timescale is 
$T_b \sim 2R  \sqrt{ r / GM(R) }$
where $r$ is the cavity radius and $R$ is the distance to the center of
the galaxy.  This assumes that the surrounding gas has to fall around
the cavity at speed $v \sim \sqrt{ GM(R) r / R^2}$ and that the
cavity must move $\sim2 r$ before gas needs to be replaced.   Our
picture is of a light bubble which is temporarily overpressured relative to
its surroundings and thus rises and expands
as the ambient density dropped.  The timescale for buoyant rise 
always exceeds that for reaching local pressure equilibrium.

We take the cavity as 1 kpc diameter at 5 kpc from nucleus. 
We derive $M(R)$ from the Galaxy mass model,
$M(R) \sim 10^{10} R_{\rm kpc} (V_{\rm max}/220)^2 M_\odot$
such that the mass goes as $R$ (in kpc) beyond 3 kpc \citep{Fr96}
assuming a spherical potential. The last term involves $V_{\rm max}$ which
is the deprojected rotation velocity on the flat part of the rotation.
curve. In Paper IV we find for NGC 3079 that 
$V_{\rm max} = 250$ km~s$^{-1}$ such that
$M(5\ {\rm kpc}) \sim 6 \times 10^{10} M_\odot$, so
$T_b \sim 1 \times 10^7$ yrs.

To form the whole superbubble, the jet
would have had to precess from its current orientation at the S corner.
However,
the four vertical bundles of filaments that constitute the superbubble and
which all break at the same
height above the disk are not explained by jet precession or bouyancy.
To inflate this structure we now examine
a wide-angle wind as proposed by DS.

\subsection{\label{sec:discuss}Dynamics of the Wind-Blown Superbubble}

As an outflow inflates a superbubble,
the various gas phases organize into a characteristic pattern of shock and ionization
fronts, as described by e.g.\ \cite{Sc85,Sc86} and simulated by e.g.\
\cite{St00}: from largest to smallest
radii in a mature superbubble there is undisturbed ISM, perhaps a photoionized
precursor, a dense shell of compressed ISM that forms a standing ``ring" shock,
the contact discontinuity between the two fluids, hot shocked wind, a wind shock,
and free-blowing wind near the sites of energy injection. Several such structures
may be evident, depending on the density structure of the galaxy ISM and on
the detailed history of the power source. The hot, shocked wind
cools by lateral expansion and by mixing with the ambient ISM at the Rayleigh-Taylor
(RT)- and Kelvin-Helmholtz-unstable interfaces between the two fluids. 

If the wind is sufficiently
energetic or long-lived to {}``breakout{}'', then its
flow becomes a strong function of polar angle seen from the nucleus: free
flow up to a critical angle wherein its ram pressure becomes comparable
to the thermal
pressure in the diffuse component of the ISM; at larger angles a standing
bow shock in the galaxy disk decelerates and deflects wind around a region
of undisturbed ISM gas.  The wind subsequently reaccelerates into
the halo. In the following subsections, we will identify many of these structures
in our data.
Table \ref{tab:dynamics} lists dynamical parameters derived in Paper II, 
scalings we will use in \S4.3.
\placetable{tab:dynamics}

\subsubsection{ISM ``Ring" Shock}

Except for the volume occupied by the superbubble filaments, the concave bowl
of a pre-blowout superbubble must be filled largely with shocked wind 
\citep[zone
B in the nomenclature of][]{We77}. The free-wind zone A is on a much smaller
scale, presumably within the cavity of the molecular ring in NGC 3079.
If the wind
has not encountered enough background gas to halt its free expansion, it
blows out and then the shocked wind is found only near the outer
(ISM) shock in a ring near the galaxy disk.

We can infer from Fig.\ \ref{fig:closeup}b that the wind must have a high enough
luminosity to have pushed the ring (ISM) shock to $ \approx 8\arcsec =900 $
pc radius near the disk plane. The models of \cite{Sc85} show that this occurs
when $ L_{w}\ga 10^{42} $ ergs~s$ ^{-1} $ for reasonable ISM properties
for NGC 3079, and the resulting flow can reach the galaxy halo if it
persists.
The smallest wind luminosity
for blowout occurs when the \emph{diffuse ISM} (i.e., neglecting dense clouds)
swept up by the wind becomes comparable to the wind mass, at time $ t^{*} $
by \cite{Sc85}. For an exponential variation of ambient gas density $ e^{-R/a} $,
we have $ v_{w}t^{*}=a $ when $ L_{w}=6.2\times 10^{43}a_{kpc}n_{-1}v_{9}^{3} $
ergs~s$ ^{-1}, $ with $ a $ the scale height
in kpc, $ n_{-1} $ the gas density in
$ 0.1 $ cm$ ^{-3}, $ and $ v_{9} $ the wind velocity in $ 10^9 $
cm~s$ ^{-1}. $ The equal-pressure stagnation/stand-off radius in the galaxy
plane $ R_{EP}=a\Omega =0.9\sqrt{L_{43}/v_{9.3}P_{-10}} $ kpc, where $ \Omega ^{2}=L_{w}/(2\pi P_{ISM}v_{w}a^{2}), $
a combination of wind and ISM parameters.
At blowout, \cite{Sc85} has argued that the contact discontinuity
between shocked wind and ambient ISM lies near the ISM ring
shock, a conjecture
supported by simulations \citep[e.g.][]{St00}.
Pressure balance then sets the angle $ \psi  $
between the radial wind direction $ R=\sqrt{r^{2}+z^{2}} $ and the surface
normal to the contact discontinuity \[
\cos ^{2}\psi (x,y)=\frac{x^{2}+y^{2}}{\Omega ^{2}}h(x,y),\]
where $ x=r/a $, $ y=z/a $, and $h(x,y)$ is the pressure distribution.
The shape of the contact discontinuity
follows by solving the quadratic equation \[
(y')^{2}[x^{2}-\frac{(x^{2}+y^{2})h(x,y)}{\Omega ^{2}}]-2xyy'+[y^{2}-\frac{(x^{2}+y^{2})f(x,y)}{\Omega ^{2}}]=0\]
 for $ y'\equiv dy/dx $ and then numerically integrating an $ \Omega  $-parameterized
model from the initial value in the disk-plane $ y(x=R_{EP})=0 $. 

\begin{figure}
\includegraphics[scale=0.35]{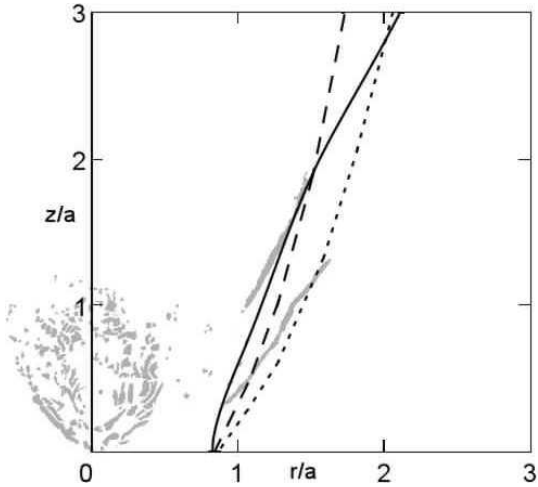}
\caption{
The shape of the contact discontinuity/ISM shock for various density
distributions of the ambient ISM is shown to scale
with the superbubble filaments.
The curves delineate the equal-pressure stagnation distance at different
hieghts above the disk. The 
dimensionless parameter \protect$ \Omega =\protect $
0.5 for all curves, reasonable for this outflow (see text). The solid line plots
the shape (see \S4.2.1) for exponential density functions
in both coordinates. The ISM density depends only on radius for the other lines:
dotted has
form \protect$ x^{\alpha }\protect $ with \protect$ \alpha =-3/2,\protect $
dashed has form \protect$ e^{\alpha x}\protect $ with \protect$ \alpha = -1.9\protect $.
\label{fig:cdshape}}
\end{figure}

We considered
several forms for $ h(x,y) $: exponentials in both $ x $ and $ y $,
exponential only in $ x $ ($ y $ constant over the height of the superbubble),
and power-law only in $ x $ (again, $ y $ constant). 
Fig.\ \ref{fig:cdshape}
scales these forms to match the observed shape of the outer shock. 
The H~I and diffuse ionized gas scale-heights
of NGC 3079 are uncertain but may be $ \approx  $0.5 kpc \citep[\S3.1 of paper II, respectively]{Ir91}
to 1 kpc, whereupon the filaments extend to $ y/a=4-2 $.
Evidently the calculation is not very constrained, but the
simulations of \cite{Su94} to be discussed shortly do match our
images with an exponential disk and spherical densities for the stellar spheroid
and halo.
\placefigure{fig:cdshape}

\subsubsection{\label{sec:blowout}Comparison to Hydrodynamical Models}

The \emph{HST} images strengthen our assertion in Paper II that the superbubble
has {}``blown out{}'' of the galaxy disk: the linear X-filaments discussed
in Papers II and III are now shown to join the galaxy disk at radii of $ \approx 8 $\arcsec, forming
the {}``concave bowl{}'' that we attribute to the standoff outer disk shock/contact-discontinuity.
Other features also suggest a disk shock: 

\begin{enumerate}
\item \cite{Sc85} estimated the evolving shape of the wind-driven complex,
see his Fig.\ 8. For wind parameters $ \log L_{w}=43,\, \log v_{w}=9, $
and $ \log P_{ISM}=-10 $ (cgs units and values similar to those inferred
by DS except for being twice their wind velocity), the inner
(wind) shock lies at two scale height radii where we see the filamentary shell, 
while the outer (ISM) ring
shock sits at $ \approx 3 $ scale heights in the disk where we see
the edges of the concave bowl.
\item The structure has a large {[}\ion{N}{2}{]}/H$ \alpha  $ flux ratio that is correlated
with line widths of 100-250 km~s$ ^{-1} $ (Paper III), consistent with shock
emission.
\item For central gas density 100 cm$ ^{-3} $ and wind speed $ v_{w}=10^{8.7} $
cm~s$ ^{-1} $, the wind model of DS has a central cavity at 0.5 kpc radius which
expands at 300 km~s$ ^{-1}. $ The radius is about half what we
observe up the bubble (but similar to the bright green region at its base
in Fig.\ \ref{fig:bubble}),
while the expansion rate they expect is comparable to what in \S\ref{sec:velmodel}
we attributed to that of the back wall of the bowl.
That gas reaches 550 km~s$ ^{-1} $ relative to v$_{sys}$. 
\cite{Ha95} rescale the parameters of the DS model 
to the higher densities expected in the molecular disk
at a radius of 50 pc, and obtain reasonable agreement with shock velocities
inferred from their H$ _{2} $ spectra.
\end{enumerate}

\cite{Su94} and \cite{St00} simulated
the appearances at X-ray and optical temperatures of a
superbubble at various stages of its wind-driven
evolution.
They parameterized models by temperatures,
scale heights, masses, and densities of the galaxy disk and isothermal halo.
Suchkov et al.\ did not include a stellar disk because the bulge gravity dominates;
their model injects energy within a radius of 150 pc,
comparable to the molecular disk in NGC 3079 (Fig.\ \ref{fig:closeup}).
They discussed the time evolutions of the temperature and density distributions.

To constrain parameters of NGC 3079 we used the X-ray results from PTV,
who inferred from the spatial variation of \emph{ROSAT} PSPC spectra (and the
HRI image) of NGC 3079 a hot halo with density $ n_{e}\sim 7\times 10^{-4}f^{-1/2} $
cm$ ^{-3} $; depending on the average value of $ f $ in the ionized filaments,
this $ n_{e} $ is consistent with either of the Suchkov et al. ``B{}''-models.
In contrast, the ``A" models have a large halo mass, and lead to
flows that differ dramatically from what is observed in NGC 3079, viz.
the relatively large mass above the disk causes the wind to breakout
laterally across the disk, not what we observe.
Mapping the putative hot halo awaits the increased
sensitivity of the \emph{Newton} \emph{X-ray Observatory} (\emph{XMM}); 
these data may provide
evidence that the halo results from extensive star formation in the
galaxy disk that predated the superbubble. 

However, with current X-ray data
we can say that the \emph{HST} emission-line image more closely resembles the predictions
of Suchkov et al. model B1 rather than B2. B1 has lower disk gas density
than B2 ($ n_{0}=2 $ vs. 12 cm$ ^{-3} $), is hotter
($ T_{disk}=6\times 10^{4} $
vs. $ 1\times 10^{4} $ K), and has ten times the halo density hence
a higher disk vertical scale-height (160 vs. 60 pc). 
The hotter gas disk in B1 seems more appropriate for a star-forming system
like NGC 3079.  The halo density of B2 is too tenuous to confine the wind.
B1 produces
a conical {}``nozzle{}'' with length/diameter ratio comparable to that of
the concave bowl in NGC 3079. In contrast, the tenuous halo of B2 is
much less collimating.

\begin{figure}
\includegraphics[scale=0.83]{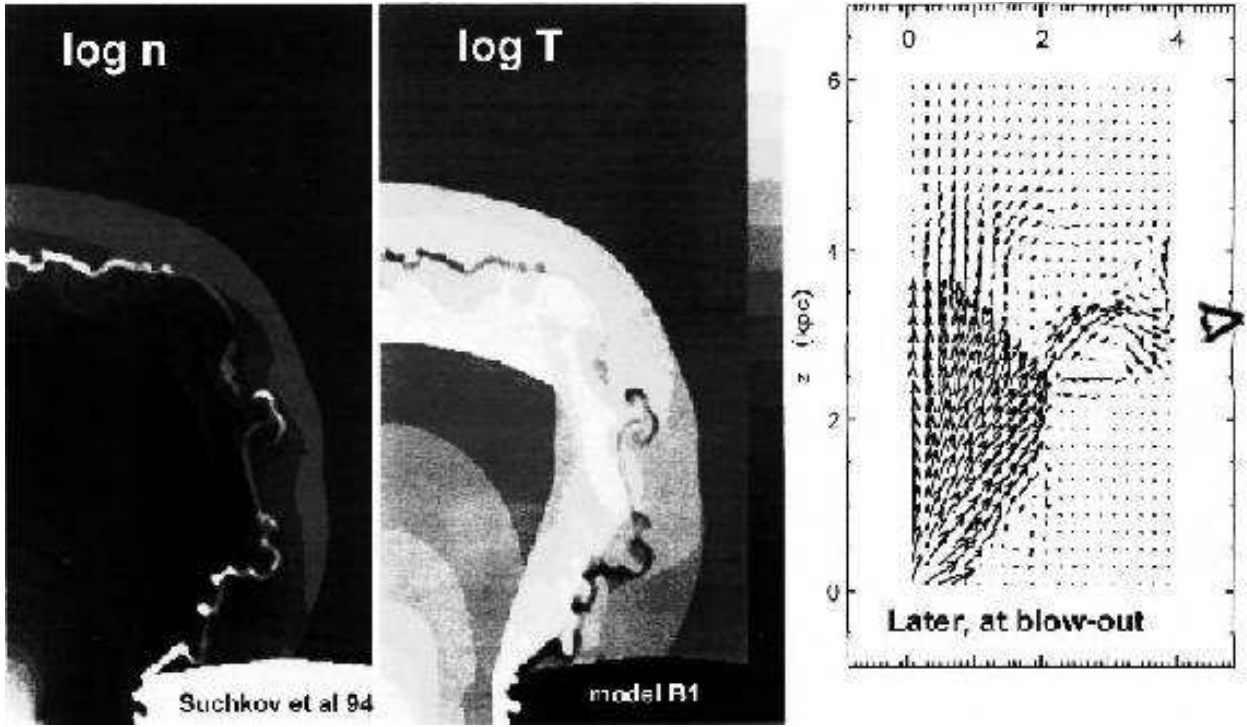}
\caption{
Densities, temperatures, and velocities from model B1
of \cite{Su94}. The first two panels are shown at time 2 Myr, the third at 3.2
Myr (i.e. after gas blowout) for a model system with more energetic parameters
than NGC 3079. For these simulations, image size is \protect$ 2\times 4\protect $
kpc\protect$ ^{2}\protect $, the range of \protect$ \log n\protect $ is
{[}-3.9, 0.8{]} in cm\protect$ ^{-3}\protect $, the range of \protect$ \log T\protect $
is {[}4,8{]} in K, and the outflow velocity reaches 3100 km~s\protect$ ^{-1}.\protect $
Note that much of the space velocity can be projected along
our line of sight (the eyeball depicts our perspective on the superbubble).
\label{fig:suchkov}}
\end{figure}

\placefigure{fig:suchkov}

Peak velocities derived for disk-gas filaments entrained in the
wind are $ \sim  $1300 km~s$ ^{-1} $ for B1 vs. $ \approx 700 $ km~s$ ^{-1} $
for B2 \citep{Su94}; these straddle observed values in the superbubble
(Paper II).
Fig.\ \ref{fig:suchkov}c shows 
velocities of B1 during breakout at 3.2 Myr (older than the superbubble
in NGC 3079).
At this stage of the model
outflow, much of the space velocity near the top of
the bubble is projected along our line of sight.
The \emph{HST} images do not suggest vortices until radii of $ \approx 10 $\arcsec.
The models predict that the vortices will eventually
propagate down the walls and grow to disrupt the superbubble completely.

Another feature of model B1
resembles the \emph{HST} structures: because the halo is so tenuous, 
entrained disk material tends to shatter into individual filaments.
Each cloud impedes the wind flow separately \textit{via} a stand-off bowshock
upstream in the wind.  The shocked wind acts to confine hence preserve
the cloud during its acceleration.  Such behaviour persists in
three-dimensional numerical simulations \citep[e.g.][]{St92}.
Overall, the B models reproduce many observed characteristics of this outflow.

\subsubsection{\label{sec:radiodiscuss}
Wind Compression of the Magnetic Field}

\begin{figure}
\includegraphics[scale=0.8]{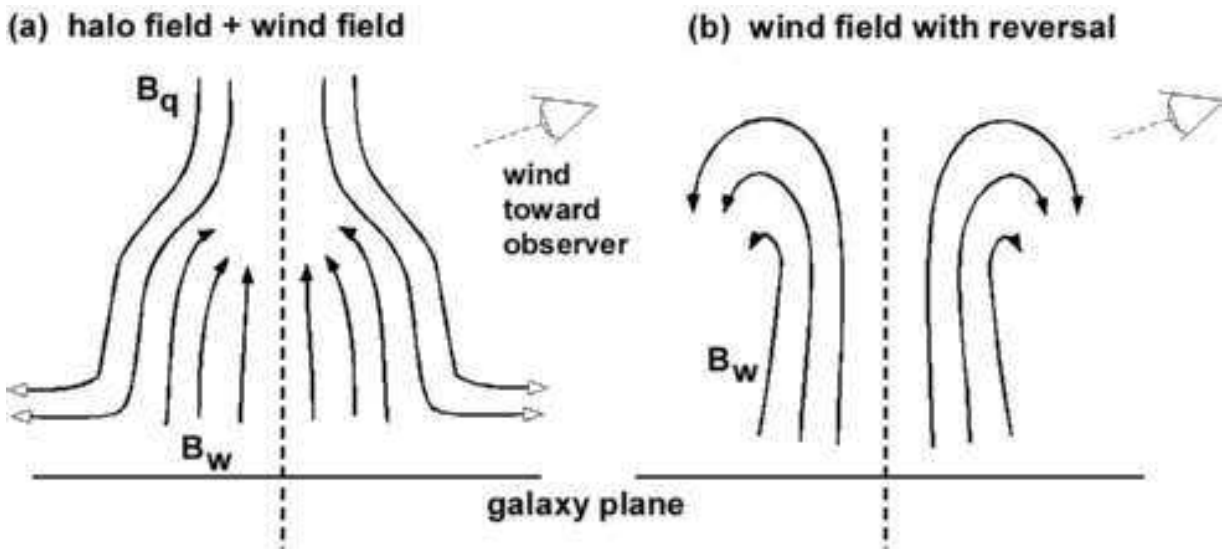}
\caption{
Two models which attempt to explain the inversion of rotation measures
by invoking a field reversal in a wind that is inclined to our line
of sight. In (a), regions of $R_m > 0$ arise from the inner surface
of the compressed radio-emitting shell. The magnetized wind ($B_w$)
expands into a diffuse magnetized halo ($B_q$) with quadrupole parity 
(and opposite
polarity). Here, regions of $R_m < 0$ arise where the wind has
compressed the halo medium. In (b), the dragged field reverses
where the entrained medium cools out of the flow. The 
outflow produces $R_m > 0$ and the returning medium reverses
the field to give $R_m < 0$.  \label{fig:bfield}}
\end{figure}

The radio data also support a blowout vortex.
The dramatic discontinuity in $R_m$, high polarization levels, and
aligned field may arise from the strong compressional shock of the 
wind. Models \citep[hereafter BMS]{BS95}
show that outflows are expected to entrain disk material and frozen-in 
magnetic field.  While initially highly turbulent, the field becomes stretched 
and compressed along the flow. Here we associate the positive $R_m$ regions 
with the updraft material, see Fig.\ \ref{fig:bfield}b.

\placefigure{fig:bfield}

BMS show that as the entrained material cools, it
rains back toward the disk outside the flow thereby
reversing the field. \textit{This behavior is seen
in the hydrodynamical simulations} discussed in \S4.2.2.  By associating
sinking gas with regions of negative $R_m$, we obtain a plausible way to reverse
$R_m$ in the E lobe (Fig.\ \ref{fig:radiopol}c).

BMS build their model on the observation that most halo 
fields appear to lie parallel to the Galactic disk \citep[e.g.][]{Be96}. 
In our interpretation of Fig.\ 3 in BMS, the halo field is not seen. When these 
models are density weighted, only the central updraft and reverse flow 
are evident.  The line of neutral magnetic field 
between regions of positive and negative $R_m$ is probably unresolved by 
our beam, and need not depolarize significantly.

An interesting variant on the BMS scheme is to associate the negative $R_m$
material with compressed, magnetized halo material (Fig.\ \ref{fig:bfield}a). 
Global magnetic fields in galaxies are often
well ordered and axisymmetric \citep{Be96}, and can
extend many kiloparsecs into the halo. Dynamo models are 
summarized by \citet[hereafter ZRS]{Ze83}.
Disk and halo fields can have odd (dipole) or even (quadrupole)
parity, and may even be of different polarity. This forms the basis for our
alternative model for the field reversal in Fig.\ \ref{fig:radiopol}c.
Again the wind drags its own field, but now runs into a quadrupole parity
halo field with polarity opposite to the wind \citep[e.g.][]{Br93}.
The sign of the rotation measure over the W lobe may establish
if we are probing the compressed halo field or
entrained material in outflows. Either outcome would be of interest.

A more exotic possibility \citep{Yu87}
is that the radio loop arises from a twisted field that protrudes from the 
disk, much like a solar prominence. This topology could explain the 
$R_m$ inversion. But, unlike a prominence, the loop cannot produce the 
observed synchrotron by winding up the field at the base of the loop.
Magnetic energy in the twisted field increases at the rate
$W \sim B^2 v_\phi A / 8\pi$, where $A$ is the footprint area and $v_\phi$
is the winding velocity. If we associate 1\% of the loop's radio energy with
the winding, for a 300~pc diameter footprint the required twist rate
is more than a hundred times larger than the vorticity or shear 
($\sim 10$ km s$^{-1}$) associated with rotating disks.
The disk field is entrained and 
dragged with the highly energetic flow.  The wind is expected to evolve
faster than the dynamo (10$^{6-7}$ yr vs.  10$^{8-9}$ yr, respectively).
The mean magnetic field in the wind seems high.
But circumnuclear activity may provide the turbulent
dynamo that feeds the azimuthal field through differential rotation. 
This can be converted to vertical field if there are azimuthal gradients
in the disk \citep{So90}.  

\subsubsection{\label{sec:survive}Disruption of Superbubble Filaments}

Our images resemble the models of \cite{Sc85}, suggesting
that the ionized filaments in the superbubble lie near the wind shock.
In detail, this shock may take the form of \emph{separate} bow shocks
upstream of each dense filament. Bathed in hot shocked gas that flows
by subsonically, the cool filaments will erode from thermo- and hydro-dynamical
instabilities. However, filaments have survived long enough for the thermalized
wind to expand for several kpc to form the stand-off shocked boundary of the
concave bowl. What does this interpretation of the data say
about the filaments?

Table \ref{tab:parms} gives densities of the ionized outer sheath of the
filaments where
column length $ N_{cloud}>N_{sheath}>3\times 10^{19} $
cm$ ^{-2} $, whereas the estimate from dust in \S\ref{sec:dust} applies
only to non-line emitting structures.
Because filaments have lofted to $ \approx  $1 kpc height, we can
estimate their minimum column to survive saturated thermal conduction 
\citep{Co77}.
To resist evaporation \citep{Kr81} requires
that $ N_{cloud}>5\times 10^{21}{\rm cm}^{-2}(R_{kpc}/v_{600})^{6/7}T^{-1/7}_{c,4}P^{6/7}_{c,9} $,
where $ P_{c,9} $ is the thermal cloud pressure in $ nT=10^{7} $ cgs units,
$ R_{kpc} $ is the height of the surviving filaments in kpc, and $ v_{600} $
is their space velocity in units of 600 km~s$ ^{-1}. $ The embedded magnetic
fields (\S\ref{sec:radiodiscuss}) are too weak to alter
these numbers much \citep{Mi93}.

\cite{Su94} show that RT instabilities grow as
$ t_{RT}=\sqrt{\lambda z\rho /2\pi v_{circ}^{2}\Delta \rho }, $
where $ \rho  $ is the combined density of the ISM gas,
and $ \Delta \rho \approx 0.01 $. These authors note that the wavelengths
of perturbations which eventually disrupt the outer shell of a superbubble are
of order the shell thickness. In NGC 3079, we see that the N shell wall is unresolved,
$<0\farcs2$ FWHM so $ \lambda <17 $ pc. From the model velocity field of the 
bar-forced
disk in Paper IV, we find a circular velocity of 100 km~s$ ^{-1} $
at the radius of the superbubble. With these numbers,
$ t_{RT}\approx 0.85 $ Myr. The dynamical time to inflate the bubble is 
somewhat greater
(Table \ref{tab:dynamics}), so it is plausible that RT instabilities
shear filaments into arches at the top of the superbubble. Because
the ISM filaments are organized into discrete streams, they appear never
to have formed a dense shell; thus, the wind can indeed break out into the halo
with ease. 

\subsection{Constraints on the Power Source}

Assuming the minimum value of 
$ f \sim 3\times 10^{-3}$ from \S\ref{sec:ke}, the power source must
be able to inject KE 
$\ga1.4\times 10^{53}$ ergs into the ionized superbubble; this
is comparable to the KE in the starburst-driven outflow of
M82 \citep{Sh98}. Because the W radio lobe probably has a similar complex
of optically obscured filaments, the KE of the ionized
outflow is comparable to that driven by the AGN in NGC 1068 
\citep[$\approx 4\times 10^{53}$ ergs,][]{Ce90}. 

We showed in Paper II that the starburst age and depletion time for gas to form
stars exceed considerably the age of the superbubble.
In support
of numerous supernova remnants \textbf{}within a nuclear starburst, \cite{Pe96}
showed that the steep/inverted-spectrum radio sources are embedded at 20 cm
in nonthermal emission that extends to 4\arcsec\ along P.A. 170\arcdeg.
\cite{Me98} showed that this structure coincides with the warped molecular disk
found by \cite{So92} to have total cool gas $ 10^{10}M_{\bigodot } $ and
$ \approx 1500\, M_{\bigodot } $ of $ H_{2}(v=1\rightarrow 0)S(1) $ gas
at its center. \cite{Br97} found that dust emission at 1.2 mm within a diameter
$\sim$10\arcsec\ is much less centrally concentrated than either the CO
or radio continuum emission; they used this emission to derive
a low $ N(H_{2})/I_{CO(2\rightarrow 1)} $ conversion ratio, and hence a more
modest mass of cold gas near the center, $\approx 2\times 10^{8}M_{\bigodot }$
(adjusted for 17.3 Mpc distance). The center appears not to contain cool
dust ($ T\leq 20 $ K), thereby resembling a low-luminosity version of an
IR ultraluminous galaxy. \cite{Br97} therefore argued that stars are forming
efficiently in this region, whereas \cite{So00} find 
an ultra-high-density molecular core and maintain that its large
differential rotation inhibits star formation.

To match $ L_{w} $ the required injection rate of KE from a
nuclear starburst would be $ \approx 3\times 10^{41}$ ergs~s$ ^{-1} $
(Table \ref{tab:dynamics}).
We had noted in Paper II that the observed mid-IR flux is consistent with
a starburst of this power, asssuming no contribution from AGN emission that
may also have been reprocessed by dust.
Using the injection rate that
we derived from our new data in \S3.9 and assuming an adiabatic
superbubble, we therefore obtained
$ 3\times 10^{41}>2.6\times 10^{54}\sqrt{f}/t_{dyn} $.
Assuming that the outflow is no larger than the maximum observed velocity of
1050 km~s$ ^{-1} $ and that our reduced estimate of the KE in the superbubble
is correct within a factor of 4, this condition is satisfied even for $f = 1$.

If radiative losses are important,
the bubble may still conserve momentum. Then the observed injection
rate $ \ga 7.4\times 10^{46}\sqrt{f}/t_{dyn} $ dyne can be compared to that
predicted from the starburst model of \cite{El89} $ \dot{p}_{*}=6\times 10^{33}L_{ir,11} $
dyne $ \approx 2\times 10^{33} $ dyne. Again, a starburst suffices (in some
cases just barely) for all values of $ f $. 

Our data constrain only the ionized outflow. Entrained neutral and
molecular material \citep{Is98,Ir92} and the wind reduce the acceptable value
of $ f $ for the ionized gas. If the higher space velocities of our Paper
II model are favored over those in this paper, then a starburst wind
will work only for very small values of $ f. $

\section{FUTURE PROSPECTS}

\subsection{\label{sec:Xrayfuture}Spectra of the Warm, Post-Shock Gas}

Its large angular size is well resolved by current X-ray telescopes, so the
superbubble in NGC 3079 offers excellent prospects to study confined hot gas
in a galaxy-wind/ISM interaction. 
Both PTV and \cite{Pa97} have modeled the PSPC spectrum obtained by \emph{ROSAT};
Ptak also 
included \emph{ASCA} data (no spatial resolution but higher spectral resolution).
However, large contamination by the relatively brighter galaxy
disk and halo made their models ambiguous. 
The \emph{Chandra X-ray Observatory} (CXO) now greatly improves
spatial resolution. 
Unfortunately, \cite{St00} have demonstrated by simulation that
the superbubble dynamics are likely to remain poorly constrained
because gas that emits soft X-rays has small
filling factor compared to the shocked wind (which at $\ga10^{8.5}$ K
is too hot to see with present instruments).

However, CXO \emph{can}
distinguish between shocks and cooling gas at various \emph{T}
near the jet. As discussed in \S4.2.2, 
hot gas likely clumps to the $ \approx 1 $\arcsec\
resolution of \emph{CXO} as it cools behind strong high-velocity ($>$400 km~s$ ^{-1} $)
shocks to 10$^4$ K.
Regions with pre-shock density $ n_{0} $ (in units of cm$ ^{-3} $)
would then be separated by up to the cooling distance of
$ 88V_{s500}^{4}/n_{0} $ pc = $ 1\times V_{s500}^{4}/n_{0} $ arcsec, where
$ V_{s500} $ is the shock velocity in units of 500 km~s$ ^{-1} $. The
distance will be somewhat smaller because most of the shock is oblique,
not normal.

\subsection{Internal Properties of the Ionized Filaments}

To better understand their
dynamics, we must constrain the ionized column
lengths, hence total masses, of the filaments. The dust calculation in \S\ref{sec:dust}
and thermal evaporation timescale estimated in \S\ref{sec:survive} are useful
first steps, but column lengths are more directly constrained with maps
of emission-line fluxes. The velocity field of ionized gas associated with
the jet must also be constrained. Integral-field
spectrometers on 8-m-class telescopes can begin this work.
Images of the filaments with higher spatial resolution may
soon be possible with a laser-assisted, adaptive optics system on these telescopes
(NGC 3079 does not have a natural guidestar).
These images would further constrain $ f $ and, together with constraints
on the column length, can exclude decisively a starburst-driven wind on
energetic grounds. 

Mid-IR spectroscopy with \emph{SIRTF} will be able to
constrain the cutoffs of the IMF, and the spatial and temporal
evolution of putative starburst activity in the nuclear region, as has been
done with the ESA \emph{Infrared Space Observatory}
for M82 by \cite{Fo00}. With photo- and shock-ionization
models such spectra could establish if filaments are pressure confined,
and whether they have sufficient dust load for efficient
wind acceleration.  With filament properties
understood, the superbubble of NGC 3079 will become a laboratory
to study the galaxy-wind driven processes thought to prevail at high-redshifts.

\section{SUMMARY}

We have analyzed VLA spectra and \emph{HST} emission-line and I-band images
of NGC 3079. Using the observed gas
distribution and recombination fluxes at high spatial resolution, we have
tightened constraints on the densities of ionized filaments that delineate the
superbubble. The \emph{HST} images and pattern of magnetic fields determined
from radio polarization and rotation measures
suggest \textit{independently} that gas at the upper boundary
is lofting out of the bubble in a vortex. Numerical simulations show that a
substantial fraction of the
space velocities of these filaments would then project
along our sightline.  We therefore reassess
the velocity field across the bubble and reduce substantially previous estimates
(including ours in Paper II) of the \emph{minimum} outflow KE and
momentum. 
We find that a bright, kinematically disturbed filament at the base of 
the superbubble and a narrow, straight plume of continuum light align with 
the putative VLBI-scale radio jet, extending the signature of the
collimated component of this outflow to a projected radius of $ \sim  $250 pc.
The rest of the superbubble is composed of four gaseous columns whose 
orientation perpendicular to the disk plane and strikingly similar appearances
are not easily explained by precessing jets.
Either starburst or AGN power sources remain viable on energetic grounds
because of uncertainties
in the flux of reprocessed infrared photons from the AGN.
But either source
will have more modest steady-state requirements based on our data and
kinematical model.

\acknowledgements{
We thank the referee for detailed comments.
J. B. H. acknowledges a DEETYA grant from the Australian government
to visit the VLA AOC, and thanks
Rick Perley for assistance with calibration of the VLA data. 
We were
supported by NASA grant GO-6674 from the Space Telescope Science Institute,
which is operated by AURA, Inc., under NASA contract NAS 5-26555. S.
V. is grateful
for partial support by a Cottrell Scholarship awarded by the
Research Corporation, NASA/LTSA grant NAG5-6547, and NSF CAREER grant AST 98-74973.
A. V. F. acknowledges the support of NASA grant NAG3-3556.}

\newpage
\onecolumn

%\begin{rotate}
%\end{rotate}

\begin{deluxetable}{lcrcc}
\tablewidth{0pc}
\tablecaption{NGC 3079 --- Basic information\label{tab:intro}}
\tablehead{\colhead{Parameter} & \colhead{Region\tablenotemark{a}} & \colhead{Flux} &
\colhead{Luminosity} & \colhead{Reference} \\
 & ($\arcsec$) & & ($L_{\sun}$) &
}
\tablenotetext{a}{``total" = value for entire galaxy, the units
of most other entries in the column are diameters in arc-seconds.}
\startdata
$B_T^0$ & total & 10.45 mag & $3.1\times 10^{10}$ & 1  \\
X-ray & core & & $3.0\times10^6$ & 2  \\
X-ray & lobes & & $\sim8\times10^6$ & 2  \\
X-ray &  3 disk sources & each & $1.6\times10^5$ & 2  \\
12 $\mu$m & total & 2.81 Jy & & 3 \\
25 $\mu$m & total & 3.54 Jy & & 3 \\
60 $\mu$m & total & 52.8 Jy & & 3 \\
100 $\mu$m & total & 96.5.8 Jy & & 3 \\
40 -- 122 $\mu$m & total & & $4.9\times 10^{10}$ & 3 \\
350 $\mu$m & $\sim80$ & 10.0$\pm$1.9 Jy & & 4 \\
450 $\mu$m & $\sim80$ & 2.1$\pm$0.9 Jy & & 4 \\
800 $\mu$m & 53 & 0.92$\pm$0.1 Jy & & 5 \\
1220 $\mu$m & total & 0.5 Jy & & 6 \\
25 $\mu$m & 11 & 1.5 Jy & & 6 \\
60 $\mu$m & 11 & 22.3 Jy & & 6 \\
100 $\mu$m & 11 & 40.8 Jy & & 6 \\
800 $\mu$m & 11 & 0.35 Jy & & 6 \\
1220 $\mu$m & 11 & 0.03 Jy & & 6 \\
CO(1-0) & 21 & 231 K km s$^{-1}$ & & 6 \\
CO(2-1) & 11& 306 K km s$^{-1}$ & & 6 \\
\hline
H I & total & & 7.1$\times 10^9$ M$_{\sun}$& 7 \\
distance & $17.3\pm 1.5 $ Mpc & & & 8 \\
pos. angle & 169$\pm4\arcdeg$ & & & 8  \\
inclination & 82$\pm4\arcdeg$ & & & 8  \\
\enddata
\tablerefs{
(1) \citet{de91};
(2) PTV; (3) \citet{Yo89}; (4) \citet{Ea89}; (5) \citet{Ha95}; (6) \citet{Br97};
(7) \citet{Ir91}, with h = 0.75; (8) Paper IV}
\end{deluxetable}

\begin{deluxetable}{lcccccc}
\tablecaption{Images of NGC 3079\label{tab:observations}}
\tablenotetext{a}{Two 50 Mhz bands were used around each central frequency.
The flux standards were 3C~48 and 3C~286 \citep{Dr87}.}
\tablenotetext{b}{Our \emph{HST} exposures were made in November 1998; archive
exposures were made in March 1999 and were reprocessed with on-the-fly
recalibrations in November 2000.}
\tablenotetext{c}{``+" indicates the exposures at two dithering points.}
\tablenotetext{d}{This image has 
low signal/noise ratio on the superbubble, so was not scrutinized.
As of November 2000 the public \emph{HST} archive also contained
a STIS UV exposure of 1320 s
at central wavelength $\lambda$147 nm (P.I.\ D.\ Calzetti).  
It spans the base of 
the superbubble and nucleus, but shows only a few bright H~II regions W of
the nucleus.}
\tablewidth{0pc}
\tablehead{
\multicolumn{3}{c}{WFPC2} & \multicolumn{2}{c}{Very Large Array} & \multicolumn{2}{c}{Time on Source at \tablenotemark{a}} \\
\colhead{Filter} & \colhead{Source \tablenotemark{b}} & \colhead{Exposures \tablenotemark{c}} & \colhead{Configuration} & \colhead{Date} & 4.5 GHz & 8.06 GHz \\
& & (s) & & & (s) & (s)
}
\startdata
F814W & 11/98 & 400, 400 & B & 1/13/92 & 6660 & 6915  \\
& Archive  & 70, 70 & C & 12/9/90 & 13110 & 14670 \\
F658N & 11/98 & 1400+2700 & & & & \\
& 11/98 & 1700+2700 & & & & \\
& Archive & 400, 400, 400 & & & & \\
F547M \tablenotemark{d} & Archive & 160, 160 & & & & \\
\enddata
\end{deluxetable}

\begin{deluxetable}{rrcrrrrrc}
%\rotate
\tabletypesize{\footnotesize}
\tablewidth{0pc}
\tablecaption{Physical Properties of the Superbubble Ionized Filaments
\label{tab:parms}}
\tablehead{
\colhead{ID} & \colhead{F$_{H\alpha}\times10^{15}$} &\colhead{Space Velocity\tablenotemark{a}} &
\colhead{L$_{H\alpha}\times 10^{-37}$} & \colhead{Volume}\tablenotemark{b} &
\colhead{$n_e$\tablenotemark{c}} & \colhead{Mass$\times 10^{-3}$ \tablenotemark{d}} & 
\colhead{KE$\times10^{-51}$ \tablenotemark{e}} & \colhead{Momentum$\times10^{-45}$ \tablenotemark{e}} \\
& \colhead{(erg~s$^{-1}$ cm$^{-2}$)} & \colhead{(km s$^{-1}$)} &
\colhead{(erg~s$^{-1}$)} & \colhead{(pc$^{-3}$)} &
\colhead{($1/\sqrt{f}$ cm$^{-3}$)} & \colhead{($\sqrt{f}$ M$_{\sun}$)} & \colhead{($\sqrt{f}$ erg)} & \colhead{($\sqrt{f}$ dyne~s)}
}
\startdata
 1 & 0.15 & 75,500 & 0.54 &    $<$2999&  14.0 [10-22] &   1.2 [0.7-1.7] &  0.07,  3.13&  0.02,  0.12 \\
 2 & 2.59 & 200,500 & 9.46 &   $<$40223&  15.9 [12-25] &  18.8 [12.3-25.5] &  7.46, 46.60&  0.75,  1.87 \\
 3 \tablenotemark{f} & 9.6 & 125,500 & 32 &  372146&   9.6 [5-14] &  150 [100-300] &  11.6, 1850&  1.8,  7.4 \\
 4 & 0.25 & -200,-500 & 0.915 &    $<$5565&  13.3 [10-20] &   2.2 [1.4-2.0] &  0.87,  5.45&  0.09,  0.22 \\
 6 & 4.56 & 125,500 & 16.83 &   82576&  14.9 [11-23] &  35.8 [22.0-48.4] &  5.56, 89.02&  0.89,  3.56 \\
 7 & 2.07 & -200,-500 & 7.56 &   $<$26102&  17.7 [14-27] &  13.5 [8.8-18.5] &  5.36, 33.51&  0.54,  1.34 \\
 9 & 1.72 & -200,-500 & 6.28 &   $<$22677&  17.3 [14-27] &  11.4 [7.0-15.8] &  4.55, 28.45&  0.46,  1.14 \\
 11 & 15.1 & -350,-500 & 56.9 &  381657&  12.5 [9-19] & 139.4 [88.0-193.7] &169.88,346.69&  9.71, 13.87 \\
 12 & 1.18 & -200,-500 & 4.45 &   $<$43214&  10.3 [8-16] &  13.1 [8.8-18.5] &  5.22, 32.63&  0.52,  1.30 \\
 13 & 2.97 & -250,-750 & 11.23 &   44074&  16.4 [14-26] &  21.0 [15.0-29.1] & 13.08,117.74&  1.05,  3.14 \\
 15 & 3.58 & 175,750 & 13.8 &   94132&  12.3 [9-19] &  33.8 [22.0-48.4] & 10.30,189.18&  1.18,  5.06 \\
 16 & 1.08 & -350,-750 & 4.18 &   $<$28241&  12.3 [10-19] &  10.2 [7.0-14.1] & 12.44, 57.12&  0.71,  1.53 \\
 19 & 1.03 & -350,-750 & 3.96 &   $<$19251&  14.5 [11-23] &   8.2 [5.3-11.4] &  9.97, 45.80&  0.57,  1.23 \\
 20 & 0.98 & -200,-750 & 3.77 &   34232&  10.7 [8-17] &  10.7 [7.0-15.0] &  4.23, 59.55&  0.42,  1.58 \\
 23 & 1.00 & 0,750 & 3.87 &   $<$30806&  11.4 [9-18] &  10.2 [7.0-14.1] &  0.00,  0.00&  0.00,  0.00 \\
 24 & 2.03 & -250,-750 & 7.81 &   $<$44501&  13.4 [10-20] &  17.4 [12.3-24.6] & 10.84, 97.53&  0.87,  2.61 \\
 25 & 0.27 & -350,-1000 & 1.05 &    $<$8556&  11.2 [9-18] &   2.8 [1.9-4.0] &  3.43, 28.03&  0.19,  0.55 \\
 26 & 1.23 & -250,-1000 & 5.10 &   99270&   7.0 [6-11] &  20.3 [13.2-29.9] & 12.64,202.27&  1.01,  4.05 \\
 28 & 1.60 & 275,1000 & 6.63 &   $<$57761&  10.5 [8-17] &  17.7 [13.2-24.6] & 13.31,176.01&  0.97,  3.52 \\
 33 & 1.18 & -25,-1000 & 4.89 &   $<$59474&   8.9 [7-15] &  15.4 [10.6-22.0] &  0.10,154.94&  0.08,  3.17 \\
 34 & 1.69 & -100,-1000 & 6.995 &   79585&   9.2 [7-15] &  21.3 [15.0-29.9] &  2.12,212.16&  0.42,  4.23 \\
 36 & 0.39 & -250,-1000 & 1.60 &   $<$16686&   9.5 [8-16] &   4.7 [3.5-7.0] &  2.90, 46.34&  0.23,  0.92 \\
 37 & 2.38 & -250,-1000 & 9.80 &  $<$162161&   7.6 [6-12] &  36.1 [26.4-52.8] & 22.43,358.90&  1.80,  7.18 \\
 38 & 0.47 & -50,-1000 & 1.96 &   $<$34658&   7.4 [6-12] &   7.5 [5.3-10.6] &  0.18, 73.95&  0.08,  1.58 \\
 40 & 0.15 & 300,1000 & 0.63 &    $<$6843&   9.3 [7-15] &   1.8 [1.3-2.6] &  1.66, 18.39&  0.11,  0.38 \\
 41 & 0.44 & -250,-1000 & 1.81 &   62899&   5.2 [4-9] &   9.7 [6.2-14.1] &  6.02, 96.34&  0.48,  1.94 \\
 42 & 1.23 & 0,1000 & 5.09 &   62039&   8.9 [7-15] &  16.1 [11.4-22.9] &  0.00,  0.00&  0.00,  0.00 \\
 43 & 0.38 & 375,1000 & 1.55 &   $<$16259&   9.5 [7-16] &   4.6 [3.1-6.2] &  6.40, 45.51&  0.34,  0.92 \\
 44 & 0.08 & 400,1000 & 0.34 &    $<$3426&   9.8 [8-16] &   1.0 [0.7-14.1] &  1.54,  9.63&  0.08,  0.20 \\
 45 & 0.87 & -150,-1000 & 3.59 &   90280&   6.1 [5-10] &  16.3 [10.6-23.8] &  3.64,161.98&  0.48,  3.23 \\
 46 & 0.59 & 100,1000 & 2.45 &   $<$29520&   9.0 [7-15] &   7.7 [5.3-11.4] &  0.77, 77.47&  0.16,  1.58 \\
 47 & 0.34 & -250,-1250 & 1.395 &   41935&   5.7 [5-9] &   7.0 [4.8-10.6] &  4.32,108.06&  0.34,  1.72 \\
  48 & 0.22 & \nodata & 0.89 & $<$27600 & 5.6 [4-9] & 4.5 [3-6.5] & \nodata  & \nodata \\\
 49 & 1.90 & -125,-1250 & 7.84 &  128789&   7.6 [6-12] &  28.8 [20.2-40.5] &  4.47,447.21&  0.71,  7.13 \\
 50 & 1.13 & -100,-1250 & 4.656 &  $<$109105&   6.4 [5-10] &  20.4 [14.1-29.1] &  2.03,317.75&  0.40,  5.06 \\
 51 & 2.54 & 500,1250 & 10.47 &  151040&   8.2 [7-14] &  36.0 [25.5-49.3] & 89.52,559.51&  3.58,  8.96 \\
 52 & 0.70 & -375,-1250 & 2.89 &   $<$32945&   9.2 [7-15] &   8.8 [5.3-13.2] & 12.32,136.84&  0.66,  2.20 \\
  53 & 0.63 & \nodata & 2.60 & $<$375970 & 2.6 [2-4.5] & 28.3 [20-40] & \nodata  & \nodata \\\
  55 & 0.30 & \nodata & 1.25 & $<$34230 & 5.9 [5-10] & 5.9 [4-8.5] & \nodata  & \nodata \\\
 58 & 0.09 & 250,1250 & 0.377 &   $<$19685&   4.3 [3-7] &   2.5 [1.8-3.5] &  1.53, 38.29&  0.12,  0.62 \\
 60 & 0.27 & -250,-1250 & 1.11 &   $<$17546&   7.7 [6-12] &   4.0 [2.6-5.7] &  2.46, 61.62&  0.19,  0.97 \\
 61 & 1.38 & 150,1250 & 5.76 &  127077&   6.6 [5-11] &  24.4 [16.7-35.2] &  5.46,379.04&  0.73,  6.09 \\
 66 & 0.20 & -375,-1250 & 0.796 &   $<$50491&   4.0 [3-6] &   5.8 [3.5-8.4] &  8.13, 90.28&  0.43,  1.44 \\
 67 & 0.42 & -400,-1250 & 1.685 &   $<$42362&   6.2 [5-10] &   7.7 [5.3-11.4] & 12.32,120.36&  0.62,  1.93 \\
 70 & 1.04 & -550,-1250 & 4.19 &  105253&   6.2 [5-10] &  19.3 [13.2-27.3] & 58.00,299.57&  2.11,  4.80 \\
 71 & 0.08 & -375,-1250 & 0.307 &   $<$22250&   3.6 [3-6] &   2.4 [1.3-3.5] &  3.33, 36.97&  0.18,  0.59 \\
 73 & 0.20 & 125,1250 & 0.822 &   31233&   5.1 [4-8] &   4.7 [2.6-7.0] &  0.72, 72.19&  0.11,  1.14 \\
  75 & 0.06 & \nodata & 0.233 & $<$26500 & 2.9 [2.5-5] & 2.3 [1.6-3.3] & \nodata  & \nodata \\\
  76 & 0.13 & \nodata & 0.519 & $<$13250 & 6.2 [5-10] & 2.4 [1.6-3.3] & \nodata  & \nodata \\\
 77 & 2.07 & -100,-1250 & 8.16 &  200671&   6.4 [5-10] &  37.2 [26.4-52.8] &  3.70,577.72&  0.74,  9.24 \\
  78 & 0.05 & \nodata & 0.20 & $<$9390 & 4.6 [3.5-7.5] & 1.3 [0.8-1.8] & \nodata  & \nodata \\\
  79 & 0.14 & \nodata & 0.553 & $<$35330 & 3.9 [3-6.5] & 4.1 [2.5-6] & \nodata  & \nodata \\\
 80 & 0.19 & -400,-1500 & 0.773 &   $<$43214&   4.2 [3-7] &   5.3 [3.5-7.9] &  8.41,118.23&  0.42,  1.58 \\
 82 & 0.78 & 200,1500 & 2.768 &   $<$47066&   8.1 [6-12] &  11.1 [7.0-15.0] &  4.41,248.09&  0.44,  3.30 \\
 83 & 0.78 & -125,-1500 & 2.778 &  103548&   5.5 [4-9] &  16.6 [10.6-22.0] &  2.57,370.17&  0.41,  4.97 \\
 84 & 1.54 & 200,1500 & 5.504 &   91993&   8.2 [6-12] &  21.9 [13.2-29.1] &  8.72,490.73&  0.87,  6.54 \\
  89 & 0.09 & \nodata & 0.313 & $<$20430 & 4.1 [3-6.5] & 2.5 [1.7-3.5] & \nodata  & \nodata \\\
  90 & 0.11 & \nodata & 0.403 & $<$21530 & 4.6 [3-7] & 2.9 [2-4] & \nodata  & \nodata \\\
 92 & 0.85 & -650,-1500 & 3.02 &   $<$69742&   6.9 [5-10] &  14.2 [8.8-19.4] & 59.56,317.16&  1.83,  4.23 \\
 93 & 0.06 & 150,1500 & 0.227 &    $<$9416&   5.2 [3-8] &   1.4 [0.9-1.9] &  0.32, 31.69&  0.04,  0.44 \\
 94 & 0.65 & 550,1750 & 2.306 &   44501&   7.6 [6-11] &   9.9 [6.2-13.2] & 29.66,300.26&  1.08,  3.45 \\
  96 & 0.12 & \nodata & 0.432 & $<$29820 & 4.0 [3-6] & 3.5 [2.4-4.5] & \nodata  & \nodata \\\
  97 & 0.14 & \nodata & 0.513 & $<$32020 & 4.2 [3-6.5] & 3.9 [2.5-5.5] & \nodata  & \nodata \\\
  98 & 0.14 & \nodata & 0.499 & $<$33130 & 4.1 [3-6] & 4.0 [2.5-5.5] & \nodata  & \nodata \\\
 99 & 0.04 & -425,-2000 & 0.154 &   $<$12834&   3.6 [3-6] &   1.4 [0.9-1.8] &  2.53, 55.95&  0.12,  0.58 \\
101 & 0.37 & -550,-2000 & 1.327 &   $<$28667&   7.2 [6-11] &   6.0 [3.5-7.9] & 18.01,238.17&  0.65,  2.37 \\
103 & 1.06 & 225,2000 & 3.779 &   92845&   6.7 [5-10] &  18.2 [10.6-24.6] &  9.17,724.79&  0.82,  7.28 \\
104 & 0.12 & 250,2000 & 0.412 &   $<$29954&   3.9 [3-6] &   3.4 [2.2-4.4] &  2.13,136.35&  0.17,  1.34 \\
 105 & 0.05 & \nodata & 0.183 & $<$18770 & 3.3 [2.5-5] & 1.8 [1.2-2.3] & \nodata  & \nodata \\\
106 & 0.12 & -625,-2000 & 0.438 &   $<$11121&   6.6 [5-10] &   2.1 [1.3-2.9] &  8.20, 84.02&  0.26,  0.85 \\
 107 & 0.04 & \nodata & 0.154 & $<$13800 & 3.5 [2.5-5.5] & 1.4 [0.9-1.9] & \nodata  & \nodata \\\
109 & 0.18 & -500,-2000 & 0.631 &   $<$19685&   6.0 [5-9] &   3.4 [2.2-4.4] &  8.54,136.63&  0.34,  1.37 \\
110 & 0.10 & -900,-2000 & 0.370 &   $<$11121&   6.0 [5-9] &   1.9 [1.3-2.6] & 15.60, 77.04&  0.34,  0.76 \\
112 & 0.31 & -625,-2000 & 1.09 &   39789&   5.6 [4-9] &   6.4 [3.5-8.8] & 24.97,255.66&  0.80,  2.56 \\
 114 & 0.03 & \nodata & 0.117 & $<$17120 & 2.7 [2-4.5] & 1.4 [0.9-1.8] & \nodata  & \nodata \\\
 116 & 0.001 & \nodata & 0.004 & $<$7730 & 0.8 [0.5-1.2] & 0.2 [0.1-0.25] & \nodata  & \nodata \\\
117 & 1.36 & -900,-2000 & 4.86 &  117234&   6.8 [5-10] &  23.2 [15.0-29.1] &187.21,924.51&  4.16,  9.25 \\
119 & 0.07 & -50,-2000 & 0.237 &   $<$13268&   4.4 [3-7] &   1.8 [1.1-2.3] &  0.04, 70.43&  0.02,  0.70 \\
 120 & 0.01 & \nodata & 0.038 & $<$26500 & 1.3 [0.9-1.9] & 1.0 [0.6-1.3] & \nodata  & \nodata \\\
123 & 0.16 & -425,-2000 & 0.585 &   $<$29520&   4.7 [3-7] &   4.0 [2.6-5.3] &  7.27,161.03&  0.34,  1.62 \\
125 & 1.41 & -250,-2250 & 5.03 &   78298&   8.4 [6-14] &  19.4 [12.3-25.5] & 12.03,974.77&  0.96,  8.64 \\
129 & 0.37 & -425,-2250 & 1.33 &   87715&   4.1 [3-6] &  10.6 [5.3-14.1] & 18.98,531.97&  0.89,  4.71 \\
 130 & 0.005 & \nodata & 0.017 & $<$18220 & 1.0 [0.7-1.5] & 0.5 [0.3-0.7] & \nodata  & \nodata \\\
131 & 0.24 & -175,-2250 & 0.867 &   62899&   3.9 [3-6] &   7.2 [4.4-9.7] &  2.20,363.81&  0.26,  3.28 \\
132 & 0.50 & -800,-2250 & 1.79 &   $<$37223&   7.3 [6-11] &   7.9 [5.3-10.6] & 50.43,398.88&  1.26,  3.54 \\
133 & 0.12 & -100,-2250 & 0.440 &   $<$20964&   4.8 [3-7] &   3.0 [1.8-4.0] &  0.30,151.53&  0.06,  1.39 \\
134 & 0.58 & -700,-2250 & 2.076 &   75733&   5.6 [4-9] &  12.2 [7.0-15.8] & 59.63,616.12&  1.71,  5.49 \\
 136 & 0.02 & \nodata & 0.088 & 142440 & 0.8 [0.6-1.3] & 3.5 [2-4.5] & \nodata  & \nodata \\\
137 & 0.10 & -800,-2250 & 0.368 &   $<$18399&   4.7 [3-8] &   2.6 [1.6-3.3] & 16.25,128.55&  0.40,  1.14 \\
 138 & 0.02 & \nodata & 0.077 & $<$13800 & 2.5 [1.9-3.6] & 1.0 [0.6-1.4] & \nodata  & \nodata \\\
141 & 0.13 & -275,-2250 & 0.459 &   50491&   3.2 [2-5] &   4.7 [2.6-6.2] &  3.51,235.14&  0.26,  2.09 \\
142 & 0.19 & -250,-2250 & 0.688 &   $<$20964&   6.0 [5-9] &   3.7 [2.5-4.8] &  2.30,186.11&  0.18,  1.66 \\
143 & 1.08 & -850,-2250 & 3.87 &   65891&   8.1 [6-12] &  15.6 [10.6-20.2] &111.96,784.51&  2.63,  6.97 \\
 144 & 0.09 & \nodata & 0.303 & $<$27050 & 3.5 [2.5-5.5] & 2.8 [2-3.5] & \nodata  & \nodata \\\
145 & 0.03 & -200,-2250 & 0.099 &   $<$20964&   2.3 [2-3] &   1.4 [0.9-1.9] &  0.56, 71.31&  0.05,  0.59 \\
 149 & 0.03 & \nodata & 0.094 & 38090 & 1.7 [1.3-2.5] & 1.8 [1.2-2.5] & \nodata  & \nodata \\\
150 & 0.14 & -500,-2250 & 0.515 &   $<$24816&   4.8 [3-7] &   3.5 [2.2-4.8] &  8.75,177.20&  0.35,  1.58 \\
151 & 1.84 & -850,-2250 & 6.56 &  101835&   8.4 [6-14] &  25.2 [16.7-32.6] &180.91,1267.6 &  4.27, 11.29 \\
152 & 0.02 & -350,-2250 & 0.081 &    $<$6417&   3.7 [3-6] &   0.7 [0.4-1.0] &  0.85, 35.29&  0.06,  0.37 \\
153 & 1.47 & 700,2250 & 5.259 &  204523&   5.3 [3-8] &  32.0 [17.6-44.0] &155.72,1608.9&  4.45, 14.29 \\
156 & 0.12 & -750,-2250 & 0.443 &   $<$12408&   6.2 [5-10] &   2.3 [1.4-3.1] & 12.80,115.20&  0.34,  1.03 \\
158 & 0.09 & -750,-2250 & 0.330 &   $<$15407&   4.9 [4-7] &   2.2 [1.4-2.9] & 12.32,110.84&  0.33,  0.98 \\
 159 & 0.53 & -700,-2250 & 1.90 & 73980 & 5.3 [4-8] & 11.6 [7-16] & \nodata  & \nodata \\\
160 & 0.04 & -750,-2500 & 0.158 &    $<$8130&   4.7 [3-7] &   1.1 [0.7-1.5] &  6.40, 71.11&  0.17,  0.56 \\
 161 & 0.004 & \nodata & 0.015 &  $<$1100 & 3.9 [3-6] & 0.1 [0.1-0.2] & \nodata  & \nodata \\\
163 & 0.198 & -450,-2500 & 0.706 &   45779&   4.1 [3-6] &   5.5 [3.5-7.0] & 11.17,344.80&  0.49,  2.74 \\
166 & 0.271 & -250,-2500 & 0.966 &   62899&   4.1 [3-6] &   7.6 [4.4-10.6] &  4.71,470.98&  0.38,  3.79 \\
167 & 2.527 & -850,-2500 & 9.02 &  117661&   9.2 [7-14] &  31.7 [17.6-44.0] &227.72,1969.9 &  5.36, 15.77 \\
 169 & 0.01 & \nodata & 0.035 & 23740 & 1.3 [1-2] & 0.9 [0.6-1.2] & \nodata  & \nodata \\\
173 & 0.089 & -750,-2500 & 0.316 &   $<$26102&   3.6 [3-6] &   2.8 [0.5-1.1] & 15.76,175.09&  0.42,  1.41 \\
 174 & 0.0003 & \nodata & 0.001 & $<$13250 & 0.3 [0.2-0.45] & 0.1 [0.1-0.2] & \nodata  & \nodata \\\
179 & 0.089 & -400,-2500 & 0.310 &   $<$29094&   3.4 [3-5] &   2.9 [2.2-3.5] &  4.62,180.54&  0.23,  1.43 \\
180 & 0.027 & -275,-2500 & 0.096 &   36369&   1.7 [1-3] &   1.8 [1.3-2.6] &  1.39,114.95&  0.11,  0.96 \\
 181 & 0.0007 & \nodata & 0.003 & $<$13250 & 0.5 [0.4-0.8] & 0.2 [0.1-0.2] & \nodata  & \nodata \\\
184 & 0.18 & -750,-2750 & 0.656 &  124938&   2.4 [1.9-4] &   8.8 [5.3-12.3] & 49.25,662.09&  1.32,  4.84 \\
186 & 0.031 & -600,-2750 & 0.111 &    $<$9843&   3.5 [3-6] &   1.1 [0.6-1.4] &  3.79, 79.52&  0.12,  0.56 \\
 188 & 0.12 & \nodata & 0.427 & $<$17120 & 5.3 [4-8] & 2.6 [1.5-4] & \nodata  & \nodata \\\
189 & 0.01 & -500,-2750 & 0.037 &   $<$24389&   1.2 [1-2] &   0.9 [0.5-1.2] &  2.19, 66.31&  0.09,  0.48 \\
190 & 0.0001 & -250,-2750 & $<$0.001 &   $<$11548&   0.1 [0-0.2] &   0.1 [0.0-0.2] &  0.05,  6.39&  0.01,  0.10 \\
192 & 0.0008 & -750,-2750 & 0.003 &   38084&   0.3 [0.1-0.4] &   0.4 [0.2-0.4] &  1.97, 26.51&  0.05,  0.19 \\
194 & 0.068 & -750,-2750 & 0.241 &   $<$10269&   5.1 [3-8] &   1.5 [1.1-2.2] &  8.37,112.56&  0.22,  0.81 \\
201 & 0.111 & -750,-2750 & 0.395 &   $<$30806&   3.7 [3-6] &   3.4 [1.8-5.3] & 19.21,258.25&  0.51,  1.87 \\
 203 & 0.0070 & \nodata & 0.025 & $<$22080 & 1.1 [0.8-1.8] & 0.7 [0.5-1] & \nodata  & \nodata \\\
205 & 0.0165 & -400,-2750 & 0.059 &   $<$18833&   1.8 [1-3] &   1.1 [0.9-1.8] &  1.68, 79.47&  0.09,  0.61 \\
206 & 0.068 & -700,-2750 & 0.241 &   $<$10269&   5.1 [3-8] &   1.5 [1.1-2.2] &  7.29,112.50&  0.21,  0.83 \\
207 & 0.021 & -500,-2750 & 0.076 &    $<$7277&   3.4 [2-5] &   0.7 [0.4-1.3] &  1.75, 52.99&  0.07,  0.39 \\
208 & 0.0007 & -400,-3000 & 0.002 &   $<$17112&   0.5 [0-1] &   0.2 [0.1-0.2] &  0.28, 15.85&  0.02,  0.13 \\
 209 & 0.0007 & \nodata & 0.003 & $<$12150 & 0.5 [0.4-0.8] & 0.2 [0.14-0.28] & \nodata  & \nodata \\\
210 & 0.027 & -600,-3000 & 0.097 &    $<$7277&   3.9 [3-6] &   0.8 [0.5-1.1] &  2.83, 70.87&  0.10,  0.48 \\
211 & 0.111 & -600,-3000 & 0.395 &   $<$30806&   3.7 [3-6] &   3.4 [1.8-5.3] & 12.29,307.24&  0.41,  2.07 \\
213 & 0.086 & -750,-3000 & 0.307 &  148475&   1.5 [1-2] &   6.6 [4.4-8.8] & 36.94,591.03&  0.99,  3.94 \\
214 & 0.071 & -500,-3250 & 0.251 &    $<$7704&   6.0 [5-9] &   1.3 [0.9-2.6] &  3.28,138.74&  0.13,  0.86 \\
216 & 0.010 & -500,-3250 & 0.036 &   $<$11982&   1.8 [1-3] &   0.6 [0.3-1.1] &  1.53, 64.72&  0.06,  0.40 \\
217 & 0.114 & -500,-3250 & 0.406 &   $<$24816&   4.3 [3-6] &   3.1 [1.8-4.4] &  7.66,323.59&  0.31,  2.00 \\
 219 & 0.005 & \nodata & 0.019 & $<$10490 & 1.4 [1.1-2.4] & 0.4 [0.2-0.8] & \nodata  & \nodata \\\
 220 & 0.001 & \nodata & 0.004 & $<$9390 & 0.7 [0.5-1.1] & 0.2 [0.1-0.4] & \nodata  & \nodata \\\
 223 & 0.002 & \nodata & 0.008 & $<$7730 & 1.1 [0.8-1.6] & 0.2 [0.1-0.4] & \nodata  & \nodata \\\
224 & 0.0001 & -400,-3500 & $<$0.001 &   $<$27381&   0.1 [0-0] &   0.1 [0.0-0.2] &  0.14, 10.78&  0.01,  0.08 \\
225 & 0.013 & -450,-3500 & 0.047 &   $<$24816&   1.5 [1-2] &   1.1 [0.5-2.1] &  2.13,128.88&  0.10,  0.75 \\
226 & 0.0001 & -450,-3500 & $<$0.001 &   $<$11548&   0.1 [0-0.2] &   0.1 [0.0-0.2] &  0.18, 10.65&  0.01,  0.07 \\
227 & 0.0016 & -400,-3500 & 0.006 &   $<$15407&   0.7 [0-1] &   0.3 [0.1-0.5] &  0.42, 32.35&  0.02,  0.15 \\
228 & 0.0001 & -400,-3500 & $<$0.001 &   $<$11548&   0.1 [0-0.2] &   0.1 [0.0-0.2] &  0.14, 10.78&  0.01,  0.08 \\
229 & 0.004 & -400,-3500 & 0.015 &   $<$18825&   0.9 [0.7-1] &   0.5 [0.3-1.1] &  0.84, 64.03&  0.04,  0.39 \\
232 & 0.0001 & -400,-3500 & $<$0.000 &    $<$6417&   0.2 [0-0.4] &   0.1 [0.0-0.2] &  0.14, 10.78&  0.01,  0.08 \\
 233 & 0.0002 & \nodata & 0.001 & $<$18770 & 0.2 [0.1-0.4] & 0.1 [0.1-0.4] & \nodata  & \nodata \\
 \enddata
%\tablecomments{}
\tablenotetext{a}{Space velocity from model 
in this paper, and paper II, respectively. Velocities were assigned from radial
velocities that were mapped from the ground at 1$\arcsec$ resolution (see Paper 
II).}
\tablenotetext{b}{$<$ indicates that WFPC2 did not resolve
one or both of the cloud dimensions. The value tabulated was established by 
fitting the intensity profiles of
the brightest filaments, or estimated by eye for the rest of the
cloud population.}
\tablenotetext{c}{Ionized gas density in units shown, 
mean value and [upper-lower] range shown.}
\tablenotetext{d}{Ionized gas mass in units shown, 
mean value and [upper-lower] range shown.}
\tablenotetext{e}{Using space velocities from this paper, and paper II, respectively.}
\tablenotetext{f}{This is the main filament associated with the jet.
Flux listed is sum of low- and high-velocity line components, see \S3.10.
Velocity is the flux-averaged value.}
\end{deluxetable}

\onecolumn
\begin{deluxetable}{lll}
\tablecaption{Derived Parameters of the Ionized Gas Associated with the Jet
\label{tab:jetparms}}
\tablewidth{0pc}
\tablehead{
\colhead{Parameter} & \colhead{Value} & \colhead{Note}}
\startdata
[\ion{N}{2}]$\lambda\lambda$6548+6583/H$\alpha$ & $3\pm0.2$ & Typical of AGN \\
Velocity FWHM & 440 km~s$^{-1}$ & Corrected for instrumental resolution \\
H$\alpha$ emission & $9.6\times 10^{-15}$ ergs~s$^{-1}$~cm$^{-2}$~arcsec$^{-2}$ & 10\% of superbubble H$\alpha$ flux \\
Ionized mass & $(1-3)\times 10^5 \sqrt{f}$ M$_\odot$ & \\
KE & $(0.5-1.5)\times 10^{53}\sqrt{f}$ ergs & \\
H recombination time & $>10^4$ yrs & \\
\enddata
\end{deluxetable}

\begin{deluxetable}{ll}
\tablecaption{Derived Parameters of the Superbubble Outflow from Paper II
\label{tab:dynamics}}
\tablewidth{0pc}
\tablehead{
\colhead{Parameter} & \colhead{Expression}}
\startdata
Timescale from velocities&
$ t_{dyn}=10^{6}R_{bubble,kpc}/V_{bubble} $ yr\\
\\*
Adiabatic expansion timescale&
$ R_{bubble}=0.6 (L_{w,43}t_{bubble,6}^{2}/n_{0})^{0.2} $ kpc\\
of a wind-inflated bubble&
$ V_{bubble}=400(L_{w,43}/n_{0}t^{2}_{bubble,6})^{0.2} $~km~s$ ^{-1} $\\
is&
$ t_{bubble}=6\times 10^{5}R_{bubble,kpc}/V_{bubble,1000} $ yr\\
&
\\
KE rate into outflow&
$ \dot{E}_{kin}\approx 3\times 10^{41}E_{kin,55}V_{bubble,1000}/R_{bubble,kpc} $
ergs~s$ ^{-1} $\\
Wind mechanical luminosity&
$ L_{w,43}=3R^{2}_{bubble,kpc}V^{3}_{bubble,1000}n_{0} $ ergs~s$ ^{-1} $\\
Input momentum rate&
$ \dot{p}_{kin}\approx 3\times 10^{32}p_{kin,46}V_{bubble,1000}/R_{bubble,kpc} $
dyne\\
Wind+ionized entrainment mass&
$ \dot{M}=10M_{bubble,7}V_{bubble,1000}/R_{bubble,kpc}\, M_{\bigodot } $
yr$ ^{-1} $\\
\enddata
\tablecomments{\protect$ L_{w,43}\protect $ is the kinetic/mechanical
luminosity of the wind in units of \protect$ 10^{43}\protect $ ergs s\protect$ ^{-1}\protect $,
\protect$ L_{ir,11}\protect $ is the IR luminosity of the starburst in units
of \protect$ 10^{11}L_{\bigodot }\protect $, 
$g_{nuc}$ is the fraction of this from the nucleus,
and \protect$ \beta \protect $
is the fraction of the bolometric luminosity of the nuclear region
that is radiated by stars assuming a Salpeter initial mass function (IMF)
that extends to \protect$ 100~M_{\bigodot }\protect $.}
\end{deluxetable}

\end{document}